\documentclass{article}

\PassOptionsToPackage{numbers}{natbib}

 \usepackage[preprint]{neurips_2026}

\usepackage[utf8]{inputenc} 
\usepackage[T1]{fontenc}    
\usepackage{hyperref}       
\usepackage{url}            
\usepackage{booktabs}       
\usepackage{amsfonts}       
\usepackage{nicefrac}       
\usepackage{microtype}      
\usepackage{xcolor}         

\usepackage[pdftex]{graphicx}
\usepackage{amsmath}
\usepackage{amssymb}
\usepackage{rotating}
\usepackage{subcaption} 
\usepackage{placeins} 
\usepackage{float} 
\usepackage{multirow}
\usepackage{colortbl}       

\usepackage{pdflscape}
\usepackage{caption}
\usepackage{placeins}

\newcommand{\dimp}[1]{{\scriptsize\textcolor{red!85!black}{#1}}}    
\newcommand{\ddeg}[1]{{\scriptsize\textcolor{blue!85!black}{#1}}}   

\title{Are Compact Rationales Free? Measuring Tile Selection Headroom in Frozen WSI-MIL}

\author{%
  \begin{minipage}{\textwidth}
    \centering
    Hyun Do Jung\textsuperscript{1}, Jungwon Choi\textsuperscript{2}, Soojung Choi\textsuperscript{3}, Yujin Oh\textsuperscript{\textdagger,4}, and Hwiyoung Kim\textsuperscript{\textdagger,5} \\[0.3em]
    \normalfont\small
    \textsuperscript{1}Department of Artificial Intelligence, Yonsei University, Seoul, South Korea \\
    \textsuperscript{2}Kim Jaechul Graduate School of AI, KAIST, Daejeon, South Korea \\
    \textsuperscript{3}Department of Integrative Medicine, College of Medicine, Yonsei University, Seoul, South Korea \\
    \textsuperscript{4}Department of Biomedical Systems Informatics, College of Medicine, Yonsei University, Seoul, South Korea \\
    \textsuperscript{5}H-Data Strategy Center, Hallym University Chuncheon Sacred Heart Hospital, Chuncheon, South Korea \\
    \textsuperscript{\textdagger}Co-corresponding authors: \texttt{yujinoh@yuhs.ac}, \texttt{hykim@hallym.or.kr}
  \end{minipage}
}

\begin{document}

\maketitle

\begin{abstract}
Whole-slide image (WSI) multiple instance learning (MIL) classifiers can achieve strong slide-level AUC while leaving the full-bag prediction opaque. Attention scores are widely reused as post-hoc explanations, but high attention can reflect aggregation preference rather than a compact, model-sufficient rationale. We study \emph{post-hoc rationale highlighting} for frozen WSI-MIL: given a trained classifier, can its slide-level prediction be recovered from a compact, output-consistent tile subset without retraining the backbone? We instantiate this question with Finding Optimal Contextual Instances (FOCI), a lightweight rationale-readout layer over a frozen MIL backbone. FOCI is trained with model-output sufficiency and exclusion objectives over keep/drop tile subsets, evaluated with an insertion-style Sequential Reveal Protocol (SRP) adapted to WSI-MIL, and summarized by the Selection Headroom Index (SHI). Across three WSI benchmarks and seven MIL backbones, FOCI reveals that compact rationales are selection-headroom dependent rather than universally available: transformer and multi-branch attention aggregators can admit compact rationales, near-minimal attention-pooling baselines enter a selection-saturation regime, and hard-selection backbones can conflict with an external readout. For TransMIL, relative to its documented CLS-proxy ranking, FOCI reduces the Minimum Sufficient K (MSK) tile count by 32--56\% across the three benchmarks, while ACMIL+FOCI attains the highest mean SHI ($+0.465$). Deletion-based perturbation and selected-only downstream evaluation provide complementary checks. These results position FOCI as a model-level interpretability and audit layer: selected tiles are not claims of clinical or pathologist-level diagnostic sufficiency, but candidate rationales that offer a compact, reviewable view of when a frozen MIL prediction can be localized to a small output-consistent subset.
\end{abstract}

\section{Introduction}
\label{sec:Introduction}
Whole-slide image (WSI) classification plays a central role in computational pathology, supporting cancer subtyping, grading, and prognosis. The dominant approach extracts patch features with a frozen foundation encoder, aggregates them through a MIL backbone, and trains on slide-level labels alone~\citep{RN8, ilse2018attention, lu2021data, chen2024towards}. This pipeline reaches competitive diagnostic accuracy across several benchmarks~\citep{dimitriou2019deep,GADERMAYR2024102337}, but the full-bag prediction remains opaque: it gives a single slide label without surfacing the tiles that support it.

\begin{figure}[t]
    \centering
    \includegraphics[width=\textwidth]{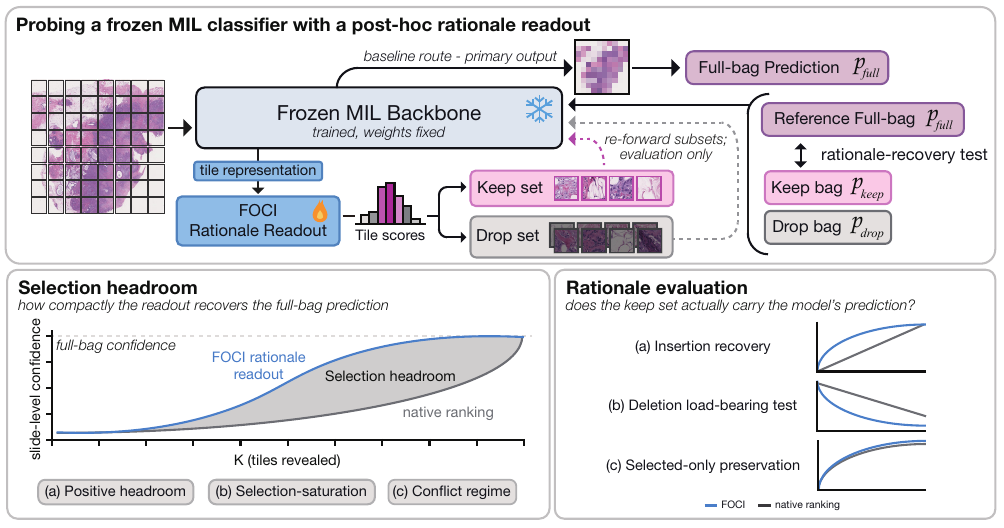}
    \caption{Selection headroom for post-hoc rationale highlighting in frozen WSI-MIL: a frozen MIL classifier produces an opaque slide-level prediction, FOCI selects a compact output-consistent tile subset that recovers it, and selection headroom across backbones determines when such compact rationales exist. On TransMIL, relative to its documented CLS-proxy ranking, FOCI reduces MSK by 32--56\% across the three benchmarks while leaving the full-bag classifier unchanged.}
    \label{fig:overview}
    \vspace{-0.75em}
\end{figure}

Attention scores are commonly repurposed as post-hoc explanations~\citep{serrano-smith-2019-attention, pruthi-etal-2020-learning, afonso2024multiple}, but high attention does not by itself answer whether a compact subset can recover the model output; in some settings, attention can reflect aggregation or training dynamics rather than an output-consistent rationale~\citep{chakraborty2017interpretability, zhu2025effective}. Such rationale highlighting may support downstream review by surfacing candidate regions for inspection, but we do not evaluate reader performance or claim clinical sufficiency.

To address this gap, we study post-hoc rationale highlighting for frozen WSI-MIL classifiers: given a trained classifier, can its slide-level prediction be recovered from a compact, output-consistent tile subset without retraining the backbone? Figure~\ref{fig:overview} summarizes this audit setting. We instantiate this question with Finding Optimal Contextual Instances (FOCI), a lightweight rationale-readout layer attached to any backbone exposing per-tile features without modifying the
existing inference pipeline.

We adapt perturbation-curve evaluation to WSI-MIL through an insertion-style \emph{Sequential Reveal Protocol} (SRP): tiles are revealed in rank order and the frozen classifier's confidence is tracked as a function of $K$. We summarize this curve with AUKC, Minimum Sufficient $K$ (MSK; the smallest $K$ that reaches $\kappa$), and Reach (fraction of slides reaching $\kappa$). SRP applies to any per-tile ranking, making it a backbone-agnostic operating-point analysis. We further introduce the \emph{Selection Headroom Index} (SHI) to quantify per-backbone compression of FOCI relative to the frozen backbone's own ranking, and we triangulate compactness with deletion-based perturbation and selected-only downstream evaluation (\S\ref{sec:experiments}).

Across three datasets---TCGA-NSCLC, TCGA-BRCA~\citep{weinstein2013cancer}, and PANDA~\citep{bulten2022artificial}---and seven MIL backbones, FOCI reveals that compact post-hoc rationales are selection-headroom dependent rather than universally available. Soft-aggregation backbones with rationale-compression headroom can be highlighted with a small tile subset, near-minimal attention-pooling baselines enter a selection-saturation regime, and hard-selection backbones can conflict with an external readout. This architecture-dependent pattern is not captured by slide-level AUC alone.

\noindent In summary, we present three contributions:
\begin{itemize}
    \item We formulate post-hoc rationale highlighting as a model-level audit layer for frozen WSI-MIL classifiers: the full-bag classifier remains unchanged, and sufficiency is used strictly in the model-output sense rather than as clinical or pathologist-level diagnostic sufficiency.

    \item We introduce \textbf{FOCI}, a lightweight rationale-readout module trained with keep/drop model-output sufficiency and exclusion objectives, and evaluate ranked subsets with SRP, MSK, AUKC, Reach, and SHI.

    \item We show that compact rationale highlighting is architecture-dependent: soft-aggregation backbones can admit compact rationales, selection-saturation regimes leave little room to improve, and hard-selection backbones can conflict with an external selector. Deletion-based perturbation and selected-only downstream evaluation provide complementary checks, with per-backbone SHI values reported in \S\ref{subsec:shi_analysis}.
\end{itemize}

Although we do not evaluate reader-level performance, the resulting candidate rationales provide a compact, reviewable view of when frozen MIL predictions can be localized to small output-consistent subsets.

\section{Related Work}
\label{sec:related}
\subsection{Multiple instance learning for whole-slide images}
The standard MIL recipe treats each slide as a bag of patch features with a single slide-level label and no per-patch annotation~\cite{dietterich1997solving, RN8}. Attention-based pooling~\cite{ilse2018attention} became the default aggregator, with CLAM~\cite{lu2021data} adding class-specific attention branches and instance-level clustering. Later WSI-MIL backbones replace or augment this pooling mechanism with transformer self-attention~\cite{shao2021transmil}, hierarchical representations~\cite{Chen_2022_CVPR}, hard instance mining~\cite{Tang_2023_ICCV}, attribution-based selection~\cite{CAI2025103631}, or multi-branch masked attention~\cite{zhang2024attention}. In parallel, frozen pathology foundation encoders such as UNI~\cite{chen2024towards}, CONCH~\cite{RN6}, and Prov-GigaPath~\cite{RN7} now provide patch features for slide-level MIL. FOCI fits this frozen-feature pipeline: the encoder and MIL backbone remain fixed, and only a lightweight rationale-readout module is trained to score which tiles to keep.

\subsection{Interpretability and faithfulness in MIL}
Attention weights are commonly reused as explanations in MIL~\cite{ilse2018attention, lu2021data,10538113}, but attention scores do not directly answer whether a compact tile subset can recover the model output~\cite{serrano-smith-2019-attention, pruthi-etal-2020-learning}. Other explanation approaches include instance-level classifiers in CLAM~\cite{lu2021data}, concept-based models~\cite{pmlr-v119-koh20a}, and gradient-based localization such as GradCAM~\cite{selvaraju2017grad}. These methods surface regions or concepts, but they typically do not report the operating-point question central to our study: how many tiles are sufficient for the frozen model to recover its prediction?

Interpretable-by-design MIL methods address related goals from a different angle. Additive MIL~\cite{javed2022additivemil} decomposes slide predictions into region-wise additive contributions, and SI-MIL~\cite{kapse2024simil} introduces a self-interpretable MIL framework with feature-level explanations. Rather than designing a new intrinsically interpretable MIL architecture, we audit frozen WSI-MIL classifiers that expose per-tile features, and we report where post-hoc rationale highlighting has selection headroom rather than claiming superiority over intrinsically interpretable models. Perturbation-based evaluation~\cite{samek2017evaluating} and MIL-specific patch-dropping metrics such as xMIL/AUPC~\cite{hense2024xmil} measure how predictions change when ranked regions are removed. Our SRP is complementary: it evaluates the insertion direction, measuring how quickly confidence is recovered as ranked tiles are progressively revealed.

\subsection{Token selection and frozen rationale readouts}
Selecting inputs that support a prediction has been studied extensively in NLP rationalization~\cite{10.3389/frai.2023.1225093}. Differentiable selectors identify token subsets that recover the full-input prediction~\cite{lei-etal-2016-rationalizing, bastings-etal-2019-interpretable, yuan2025boosting}. Related work on selective prediction~\cite{NIPS2017_4a8423d5}, early-exit networks~\cite{teerapittayanon2016branchynet}, and cooperative rationalization~\cite{yu-etal-2019-rethinking} studies adjacent questions of confidence, computation, and complement control under different training assumptions. In vision and MIL, straight-through estimators~\cite{bengio2013estimatingpropagatinggradientsstochastic} enable hard sparse selection, and ASMIL~\cite{ye2026asmil} jointly trains a selector-like mechanism with the MIL backbone.

FOCI differs from joint selector-training approaches in its frozen setting. The MIL classifier is already trained and remains fixed; the selector is a post-hoc readout head over a stable feature space, trained with keep/drop sufficiency and exclusion losses and evaluated through insertion-style SRP. This also separates FOCI from ReaMIL~\cite{jung2026reamil}, a concurrent evidence-aware MIL training method in whole-slide histopathology. In ReaMIL, the selector and backbone share gradient flow after warmup to train a compact-rationale classifier. FOCI does not train a new evidence-aware classifier. It asks whether the decisions of already-trained MIL backbones are post-hoc readable from compact tile subsets, and uses this readout to measure selection headroom and architecture-dependent failure modes.

\section{Method}
\label{sec:method}

\subsection{Frozen WSI-MIL setup}
\label{subsec:setup}

Following standard weakly supervised MIL, each slide $s$ is a bag of patch features $X_s = \{x_{s,i}\}_{i=1}^{N_s}$ extracted by a frozen encoder, together with spatial coordinates $C_s = \{c_{s,i}\}_{i=1}^{N_s}$ where $c_{s,i} = (u_{s,i}, v_{s,i})$ is the pixel location of patch $i$. We use UNI2-h~\cite{chen2024towards} to extract $d{=}1536$-dimensional features. The slide has a single label $y_s \in \{1,\ldots,C\}$ with no patch-level supervision.

Patch features are projected into a shared token space and processed by the MIL backbone (e.g., TransMIL with a learned \texttt{[CLS]} token through $L$ transformer layers); the final representation maps to class logits $\ell_s \in \mathbb{R}^C$. Full implementation and backbone details are provided in Appendix~\ref{app:setup_full}.

\paragraph{Frozen backbone.}
\label{subsec:frozen}
The backbone remains fully frozen during FOCI training; rationale losses update only the lightweight selection head ($\sim$130K parameters, under 1\% of the primary TransMIL pipeline). Joint training conflicts with the classification objective and collapses validation AUC by more than 15 points within two epochs (see Appendix~\ref{subsec:ablation}).

\subsection{Output-consistent rationale selection}
\label{subsec:problem}

Given the frozen slide classifier $f$ above, mapping a bag of tile features $X = \{x_i\}_{i=1}^{N}$ to a class probability $p$ for target class $y$, we seek a binary mask $z \in \{0,1\}^N$ with $\|z\|_0 = K$ satisfying two output-consistency conditions:
\begin{equation}
    p_y\bigl(f(z \odot X)\bigr) \ge \tau \quad \text{(sufficiency)}, \qquad
    p_y\bigl(f((1-z) \odot X)\bigr) \le \beta \quad \text{(exclusion)},
    \label{eq:problem}
\end{equation}
where $\tau, \beta$ are confidence thresholds. We call any such $K$-tile subset a \emph{model-sufficient rationale}.

\begin{figure}[t]
    \centering
    \includegraphics[width=\textwidth]{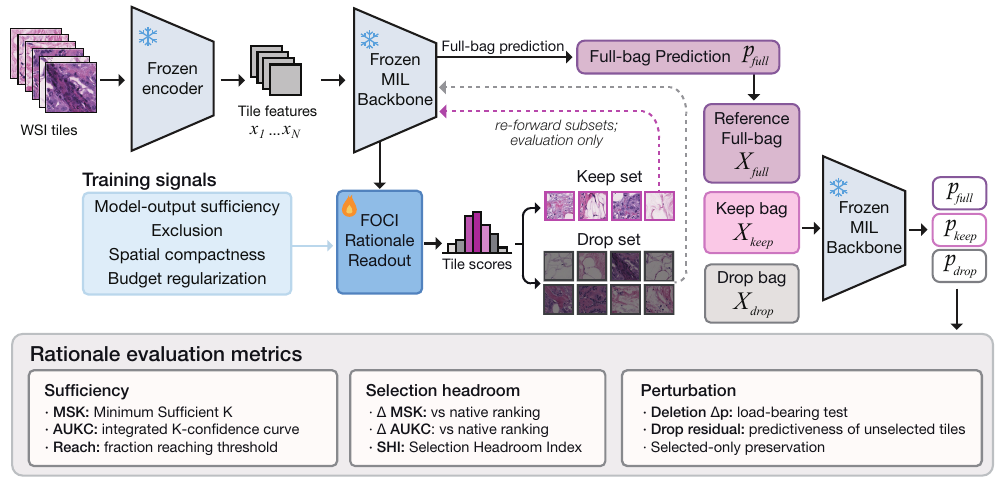}
    \caption{FOCI as a frozen rationale-readout probe. The frozen encoder maps WSI tiles to features $x_1,\ldots,x_N$, and the frozen MIL backbone produces the primary full-bag prediction. Only the lightweight FOCI selector trains; keep/drop subsets are re-forwarded through the same frozen backbone for training and evaluation.}
    \label{fig:pipeline}
    \vspace{-0.75em}
\end{figure}

\paragraph{Pipeline preservation.}
The frozen backbone $f$ continues to produce the primary slide-level prediction unchanged. FOCI does not replace the full-bag forward pass, retrain the backbone, or require pathologist annotation; it learns a per-tile scoring head that partitions each slide into a keep set (candidate rationale) and a drop set (complement). During training, the keep, drop, and full-bag views pass through the same frozen backbone with separate loss terms; at test time, tiles are ranked by the selector score and evaluated under SRP. Figure~\ref{fig:pipeline} shows the architecture.

\subsection{Rationale selection module}
\label{subsec:selector}

Given the token representations from Section~\ref{subsec:setup}, a small MLP computes a scalar selection logit $a_{s,i} = \text{MLP}_{\text{sel}}(t_{s,i}) \in \mathbb{R}$ for each token $t_{s,i}$. We consider two variants for turning these logits into selection decisions.

\paragraph{Soft gate (FOCI-Soft).}
The first variant applies the Concrete (Gumbel--sigmoid) relaxation~\cite{maddison2017the, jang2017categorical}. Sampling $\epsilon_{s,i} \sim \text{Uniform}(0,1)$:
\begin{equation}
    z_{s,i} = \sigma\!\left(
        \frac{a_{s,i} + \log \epsilon_{s,i} - \log(1-\epsilon_{s,i})}{T}
    \right), \label{eq:gumbel}
\end{equation}
where $T > 0$ is temperature. The scores $z_{s,i} \in (0,1)$ approach binary values as $T \to 0$.

\paragraph{Hard top-$K$ with straight-through (FOCI-STE).}
FOCI-STE replaces the soft Concrete gate with an exactly $K$-sparse binary mask in the forward pass while routing the backward gradient through a sigmoid surrogate~\cite{bengio2013estimatingpropagatinggradientsstochastic}, eliminating the soft-vs-hard cardinality mismatch between training and SRP evaluation. Although hard top-$K$ fixes the forward-pass cardinality, we retain a small per-bag budget/scale regularizer ($\lambda_{\mathrm{budget}} = 5 \times 10^{-3}$) to stabilize selector scores near the rank-$K$ boundary. FOCI-STE is one of two parameterizations of the same audit framework (the other being FOCI-Soft); the central object of study is whether the frozen classifier exhibits selection headroom under a consistent ranking, not the choice of gate parameterization. Full STE derivation, surrogate-gradient mechanics, and forward/backward analysis are in Appendix~\ref{app:ste_details}.

\paragraph{Three-view inference.}
Both variants produce the same three views of the slide, namely the original bag $X_{\text{full}} = X_s$, the keep bag $X_{\text{keep}} = z_s \odot X_s$ (or $m_s \odot X_s$ for FOCI-STE), and the drop bag $X_{\text{drop}} = (1 - z_s) \odot X_s$. In FOCI-Soft, all tokens stay in the sequence but non-selected patches are down-weighted by $z_s$, whereas in FOCI-STE the mask is binary. Each view passes through the frozen backbone to produce logits $\ell_{\text{full}}$, $\ell_{\text{keep}}$, and $\ell_{\text{drop}}$.

\subsection{Rationale-aware training objectives}
\label{subsec:losses}

In addition to slide-level accuracy, we design a training objective that explicitly shapes how the selector partitions tiles into a rationale subset within each bag. We compute the full-bag cross-entropy only as a preservation monitor since the full-bag forward pass bypasses the selector and the backbone is frozen. The selector itself is optimized only through losses on the keep/drop views, each enforcing a distinct property of the selection: (i) \emph{sufficiency}, where the selected tiles alone support a high-confidence prediction; (ii) \emph{exclusion}, where the remaining tiles do not support the true class; (iii) \emph{spatial compactness}, where selected tiles form a coherent region on the slide rather than scattering across it; and (iv) a small \emph{budget/scale regularizer} that controls selection mass in FOCI-Soft and stabilizes selector scores in FOCI-STE.

Let $p_y(\ell) = \mathrm{softmax}(\ell)[y_s]$ denote the true-class probability. We first define one full-bag preservation monitor and four keep/drop selector losses:
\begin{align}
    \mathcal{L}_{\text{full}}   &= \mathrm{CE}(\ell_{\text{full}},\, y_s), \\
    \mathcal{L}_{\text{suff}}   &= \mathrm{CE}(\ell_{\text{keep}},\, y_s), \label{eq:suff}\\
    \mathcal{L}_{\text{hinge}}  &= \max\!\big(\tau - p_y(\ell_{\text{keep}}),\, 0\big), \\
    \mathcal{L}_{\text{excl}}   &= \max\!\big(p_y(\ell_{\text{drop}}) - \beta,\, 0\big), \\
    \mathcal{L}_{\text{contig}} &=
        \frac{\sum_i z_{s,i}\, \lVert c_{s,i} - \mu_s \rVert_2^2}{\sum_i z_{s,i}},
\end{align}
where $\mu_s = \sum_i z_{s,i} c_{s,i} / \sum_i z_{s,i}$ is the selection-weighted centroid, $\tau \in (0,1)$ is the target confidence for the keep bag, and $\beta \in (0,1)$ is the tolerance for the drop bag. We separate the keep-bag CE term ($\mathcal{L}_{\text{suff}}$) from the confidence hinge ($\mathcal{L}_{\text{hinge}}$) because they receive different weights in the total loss.

Because the full-bag forward pass bypasses the selector and the backbone is frozen, $\mathcal{L}_{\text{full}}$ contributes no gradient to the selector parameters; we monitor it as a preservation check rather than as a selector training term. The selector objective excludes $\mathcal{L}_{\text{full}}$ and uses the keep/drop terms plus the budget regularizer:
\begin{equation}
    \mathcal{L}_{\text{selector}} =
      \lambda_{\text{suff}}\, \mathcal{L}_{\text{suff}}
      + \lambda_{\text{hinge}}\, \mathcal{L}_{\text{hinge}}
      + \lambda_{\text{excl}}\, \mathcal{L}_{\text{excl}}
      + \lambda_{\text{contig}}\, \mathcal{L}_{\text{contig}}
      + \lambda_{\text{budget}}\, \mathcal{L}_{\text{budget}}.
\end{equation}
where $\mathcal{L}_{\mathrm{budget}}$ is a small per-bag budget/scale regularizer ($\lambda_{\mathrm{budget}} = 5 \times 10^{-3}$). Full details for $\mathcal{L}_{\mathrm{budget}}$, the FOCI-Soft entropy term, and the ``sufficiency objective'' shorthand are provided in Appendix~\ref{app:loss_details}.

\paragraph{Contiguity caveat.}
$\mathcal{L}_{\mathrm{contig}}$ is a small-weighted optimization stabilizer against scattered masks, not a clinical or morphological prior. Its ablation reduces training stability (Appendix~\ref{subsec:ablation}), so we interpret it as part of the selector parameterization rather than evidence that diagnostic tissue must be spatially contiguous.

\subsection{Sequential Reveal Protocol and rationale metrics}
\label{subsec:srp}

Standard metrics such as AUC, accuracy, and F1 evaluate whether a model predicts the correct slide label, but they do not capture how much of the slide the model needed to see. Two models with identical AUC may differ in rationale compactness if one requires hundreds of tiles while the other needs only a handful.

To quantify this gap, the \textbf{Sequential Reveal Protocol (SRP)} ranks tiles by a per-tile score ($a_{s,i}$ for FOCI, or the method-specific native/proxy ranking score in Appendix~\ref{app:ranking_scores}) and reveals them in descending order; after each tile is added we record the true-class probability $p_y(K)$ to trace a confidence--count K-curve. SRP is an insertion-style adaptation of perturbation-curve evaluation~\citep{petsiuk2018rise} for WSI-MIL, with three complementary operating-point summaries (MSK, Reach, AUKC) reported per slide. Cross-method SRP reveal curves are shown in Appendix~\ref{app:srp_curves}.

\paragraph{SRP metrics.}
For each slide $s$, we summarize the $K$-curve $p_{y_s}^{(s)}(K)$ with three metrics at operating confidence $\kappa$. MSK and Reach use the threshold $\kappa$:
\begin{align}
    \mathrm{MSK}_s(\kappa) &=
    \min\bigl\{
    K \le K_{\max} \,:\,
    \arg\max_c p_c^{(s)}(K) = y_s
    \,\wedge\,
    p_{y_s}^{(s)}(K) \ge \kappa
    \bigr\}, \\
    \mathrm{Reach}(\kappa) &=
    |S_{\text{test}}|^{-1}
    \sum_{s\in S_{\text{test}}}
    \mathbf{1}\bigl[\mathrm{MSK}_s(\kappa)\text{ exists}\bigr].
\end{align}
AUKC is unthresholded. Let $K_{s,j}$ denote the $j$-th reveal count evaluated for slide $s$, let $m_s$ be the number of reveal steps, and let
$\rho_{s,j}=K_{s,j}/N_s^{\mathrm{real}}$ be the normalized reveal fraction, where $N_s^{\mathrm{real}}$ is the unpadded number of candidate tiles. We compute AUKC as the trapezoidal area under the confidence curve, normalized by the maximum reveal fraction:
\begin{align}
    \mathrm{AUKC}
    &=
    |S_{\text{test}}|^{-1}
    \sum_{s\in S_{\text{test}}}
    \frac{1}{\rho_{s,m_s}}
    \sum_{j=1}^{m_s-1}
    \frac{
    p_{y_s}^{(s)}(K_{s,j})
    +
    p_{y_s}^{(s)}(K_{s,j+1})
    }{2}
    \bigl(\rho_{s,j+1}-\rho_{s,j}\bigr).
\end{align}
$\mathrm{MSK}_\mathrm{cond}$ denotes the conditional mean over slides reaching $\kappa$. Thus, AUKC summarizes the full reveal curve, while MSK and Reach depend on the operating threshold. The three metrics expose different failure modes: AUKC can be high while Reach is low (failure on hard slides) or MSK is high (inefficient compression). We report all three at $\kappa = 0.9$ unless stated otherwise.

\paragraph{Selection Headroom Index (SHI).}
A frozen MIL classifier already produces a tile ranking through its internal
attention or aggregation weights, which itself induces a $K$-curve. To quantify
how much further FOCI compresses the sufficient set relative to that internal
ranking, we define the \emph{Selection Headroom Index}:
\begin{equation}
    \mathrm{SHI}(f) =
    \frac{\mathrm{MSK}_{\mathrm{base}}(f) - \mathrm{MSK}_{\mathrm{FOCI}}(f)}
         {\mathrm{MSK}_{\mathrm{base}}(f) + \epsilon},
    \label{eq:shi}
\end{equation}
where $\mathrm{MSK}_{\mathrm{base}}(f)$ is computed by ranking tiles with the
frozen backbone's own attention or aggregation weights,
$\mathrm{MSK}_{\mathrm{FOCI}}(f)$ uses a FOCI-trained selector attached to the
same frozen $f$, and $\epsilon$ is a small stabilizing constant. We read the
sign directly: positive SHI indicates that FOCI compresses the rationale
beyond what the backbone's internal ranking already provides, near-zero SHI
indicates a selection-saturation regime where the backbone ranking is already near-minimal, and
negative SHI indicates that the external selector conflicts with the
backbone's internal ranking enough to inflate the sufficient set. Unlike AUC,
which measures slide-level discrimination, SHI measures whether the frozen
decision can be compressed into a smaller sufficient subset, and is therefore
a property of the trained model and feature encoder rather than of
classification performance. We report SHI alongside MSK and AUKC in \S\ref{sec:experiments}. For backbones without an explicit native tile score, SHI is computed relative to a documented proxy ranking (Appendix~\ref{app:ranking_scores}) and measures improvement over the available backbone ranking, not a ranking-independent property. Our main SRP uses the ground-truth class to jointly assess confidence recovery and correctness on labeled test sets; an audit-time variant tracks the predicted class $\hat{y}=\arg\max_c f_c(X)$ instead of $y$, reported in Appendix~\ref{app:predicted_class_srp}.

\section{Experiments}
\label{sec:experiments}
We evaluate FOCI on three public benchmarks along three axes: compact rationale recovery while preserving the frozen full-bag classifier, consistency across backbone architectures and failure modes, and the contribution of each loss component (with FOCI-Soft vs FOCI-STE in Appendix~\ref{subsec:ablation}). Primary compactness results use the SRP metrics defined in Section~\ref{subsec:srp}; deletion-based perturbation and selected-only AUC provide complementary checks.

\paragraph{Hypotheses tested.}
We test four linked hypotheses:
\textbf{(H1)} slide-level AUC decouples from selection headroom;
\textbf{(H2)} non-minimal baseline MSK leaves room for FOCI compression;
\textbf{(H3)} deletion and selected-only AUC provide complementary checks rather than interchangeable explanation metrics; and
\textbf{(H4)} FOCI-STE mainly improves hard-cardinality alignment rather than serving as the central contribution.
The corresponding evidence is reported in \S\ref{subsec:shi_analysis}, Table~\ref{tab:shi}, \S\ref{subsec:faithfulness}, and Appendix~\ref{subsec:ablation}.

\subsection{Setup}
\label{subsec:setup_exp}

\paragraph{Datasets.} TCGA-NSCLC (LUAD vs.\ LUSC, 1{,}043 slides) and TCGA-BRCA (IDC vs.\ other subtypes, 1{,}126 slides) from TCGA/GDC~\cite{weinstein2013cancer}, and PANDA (benign vs. malignant, ISUP grade $\geq 1$, 10{,}615 slides)~\cite{bulten2022artificial}. We use 70/15/15 train/val/test splits matching the appendix counts; full split details and per-class counts are in Appendix~\ref{app:setup_full}.

\paragraph{Features and backbones.} Patches are extracted at 20$\times$, embedded with frozen UNI2-h~\cite{chen2024towards} ($d{=}1536$). The primary FOCI-STE backbone is a 4-layer TransMIL ($d{=}512$, 8 heads) pretrained for 20 epochs with cross-entropy. For cross-backbone experiments, FOCI is additionally applied to ABMIL, CLAM-SB, AttriMIL, ACMIL, ASMIL, and MHIM-MIL, each pretrained independently with its original objective; in all cases the backbone is frozen during FOCI training. Full architecture, optimizer, and hyperparameter settings are in Appendix~\ref{app:setup_full}.

\paragraph{Baselines and SRP scores.} For each backbone, SRP uses its native attention or aggregation logits as the ranking score; FOCI ranks tiles by its own selector head. TransMIL has no native attention head, so we use the post-encoder CLS-dot-product score as a documented proxy ranking (Appendix~\ref{app:ranking_scores}); SHI for TransMIL therefore measures improvement over this proxy.

\subsection{Main results}
\label{subsec:main}

Per-dataset SRP results for all seven backbones with and without FOCI are reported in Appendix~\ref{app:main_tables} (Tables~\ref{tab:main_nsclc}--\ref{tab:main_panda}). Across these tables, FOCI reduces MSK when the frozen backbone has rationale-compression headroom and inflates MSK when the native ranking is already near-minimal or conflicts with an external selector. A paired Wilcoxon test on the nine TransMIL (dataset, seed) observations confirms a significant MSK reduction ($p{=}0.008$, median $\Delta\mathrm{MSK}{=}{-}4.14$) but no significant AUKC change ($p{=}0.13$); per-dataset tests are underpowered ($n{=}3$), so we use the appendix tables as direction-of-effect summaries.

\subsection{Selection headroom analysis}
\label{subsec:shi_analysis}

To quantify the per-backbone effect of attaching FOCI to a frozen MIL classifier, we compute the Selection Headroom Index (SHI, defined in \S\ref{subsec:srp}) for every (backbone, dataset) pair. Table~\ref{tab:shi} summarizes per-dataset and mean SHI for each backbone family. The raw baseline MSK, FOCI MSK, $\Delta\text{MSK}$, Reach, and AUKC values are reported in Appendix~\ref{app:main_tables} (Tables~\ref{tab:main_nsclc}--\ref{tab:main_panda}). All values are 3-seed means at $\kappa = 0.9$. SHI should be read as a signed normalized effect size rather than an absolute ranking of rationale quality: when the baseline MSK is already near one tile, small absolute MSK changes can produce large negative ratios. We therefore interpret SHI together with the raw MSK and $\Delta$MSK values in Appendix~\ref{app:main_tables}, using it to identify headroom, saturation, and conflict regimes rather than to rank backbones in isolation.

\begin{table}[t]
\centering
\caption{Selection Headroom Index (SHI) per backbone and its per-dataset breakdown. SHI is normalized by the baseline Minimum Sufficient K (MSK) tile count, so extreme values may occur when the baseline MSK is small; see Appendix~\ref{app:main_tables} for raw MSK.}
\label{tab:shi}
\resizebox{0.85\columnwidth}{!}{%
\begin{tabular}{llcccc}
\toprule
\multirow{2}{*}{Family} & \multirow{2}{*}{Backbone} & \multicolumn{4}{c}{Dataset} \\
 & & NSCLC & BRCA & PANDA & Mean \\
\midrule
\rowcolor{green!8}
\multicolumn{6}{l}{\textit{Soft-aggregation (positive headroom)}} \\[2pt]
\quad Transformer & TransMIL & \dimp{$+0.562$} & \dimp{$+0.317$} & \dimp{$+0.357$} & \dimp{$+0.412$} \\
\quad Multi-branch attention & ACMIL & \dimp{$+0.705$} & \dimp{$+0.337$} & \dimp{$+0.354$} & \dimp{$\mathbf{+0.465}$} \\
\addlinespace
\rowcolor{yellow!10}
\multicolumn{6}{l}{\textit{Attention pooling (dataset-dependent saturation)}} \\[2pt]
\quad Attention pool & ABMIL & \dimp{$+0.425$} & \ddeg{$-1.914$} & \ddeg{$-0.319$} & \ddeg{$-0.603$} \\
\quad Attention pool & CLAM-SB & \dimp{$+0.248$} & \ddeg{$-2.542$} & \dimp{$+0.016$} & \ddeg{$-0.760$} \\
\quad Attribution-based & AttriMIL & \dimp{$+0.327$} & \ddeg{$-0.931$} & \ddeg{$-0.336$} & \ddeg{$\mathbf{-0.313}$} \\
\addlinespace
\rowcolor{red!8}
\multicolumn{6}{l}{\textit{Hard selection (architectural conflict)}} \\[2pt]
\quad Hard selection & ASMIL & \ddeg{$-2.055$} & \dimp{$+0.191$} & \ddeg{$-1.441$} & \ddeg{$-1.102$} \\
\quad Hard instance mining & MHIM-MIL & \ddeg{$-0.309$} & \ddeg{$-0.229$} & \ddeg{$-0.179$} & \ddeg{$\mathbf{-0.239}$} \\
\bottomrule
\multicolumn{6}{l}{\scriptsize\textit{Note.} {$\mathrm{SHI}>0$}: FOCI compresses beyond the baseline ranking, $\mathrm{SHI}\approx0$: little room to compress, $\mathrm{SHI}<0$: FOCI} \\
\multicolumn{6}{l}{\scriptsize conflicts with the backbone's native selection; \textbf{bold}: best mean SHI within family.}\\
\end{tabular}%
}
\end{table}

Three patterns emerge: (1) TransMIL and ACMIL show consistently positive SHI across all three datasets ($+0.32$ to $+0.71$); (2) attention-pooling backbones (ABMIL, CLAM-SB, AttriMIL) show a dataset-dependent saturation regime, improving NSCLC baselines but inflating the sufficient set on BRCA, on which baseline MSK is already near-minimal ($\approx 1.1$ for ABMIL/CLAM-SB); and (3) hard-selection backbones (ASMIL, MHIM-MIL) mostly inflate under FOCI, which reflects architectural conflict between native instance selection and an external selector.

\begin{figure}[t]
    \centering
    \includegraphics[width=0.62\linewidth]{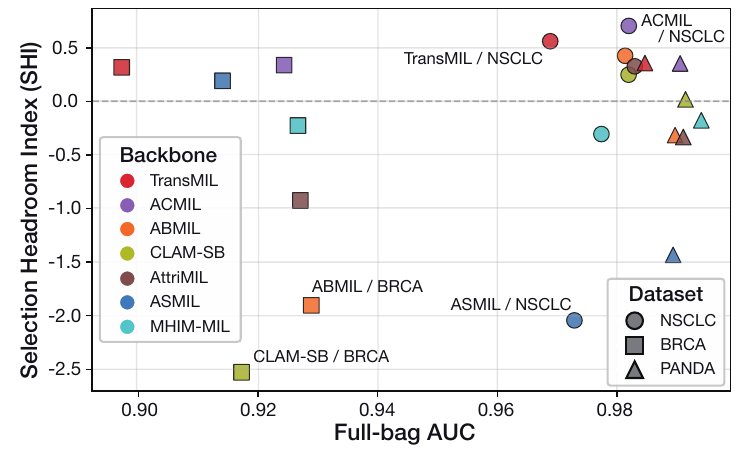}
    \caption{Slide-level AUC and SHI are decoupled. Each point is a (backbone, dataset) pair; color denotes backbone and marker shape denotes dataset. High-AUC backbones can have near-zero or negative SHI when their native ranking saturates or conflicts with an external readout, while TransMIL and ACMIL maintain positive SHI without necessarily being the best full-bag classifiers.}
    \label{fig:auc_vs_shi}
\end{figure}

Figure~\ref{fig:auc_vs_shi} visualizes this decoupling: selection headroom is not predicted by slide-level AUC alone. Predicted-class SRP (Appendix~\ref{app:predicted_class_srp}) follows the same qualitative MSK-compression pattern on TransMIL$\pm$FOCI, which supports the interpretation that the readout recovers the frozen model's own decision rather than exploiting label-specific evaluation.

\subsection{Selected-only downstream triangulation}
\label{subsec:selected_only_main}

A complementary check asks whether the compact rationale preserves the downstream TransMIL prediction. Table~\ref{tab:selected_only_compact} reports full-bag AUC and selected-only test AUC for random $K{=}32$, FOCI fixed-$K{=}32$, and FOCI adaptive-$K$ within the same TransMIL pipeline; full per-seed, per-$K$, and ABMIL-pipeline reference results are in Appendix~\ref{subsec:selected_only}.

\begin{table}[h]
\centering
\caption{Selected-only downstream AUC within the TransMIL predictor pipeline (3-seed mean). Random and fixed-FOCI rows use $K{=}32$; FOCI adaptive uses $K_s{ = }\max(16, \lfloor 0.03 N_s \rfloor)$. ABMIL native top-$K$ comparison and full per-seed / per-$K$ results are in Appendix~\ref{subsec:selected_only}.}
\label{tab:selected_only_compact}
\small
\setlength{\tabcolsep}{4pt}
\begin{tabular}{lcccc}
\toprule
Dataset & Full bag & Random $K{=}32$ & FOCI $K{=}32$ & FOCI adaptive $K_s$ \\
\midrule
NSCLC & $0.974$ & $0.969$ & $0.954$ & $0.963$ \\
BRCA  & $0.907$ & $0.881$ & $0.885$ & $\mathbf{0.907}$ \\
PANDA & $0.989$ & $0.931$ & $0.945$ & $0.934$ \\
\bottomrule
\end{tabular}
\end{table}

This mini-table is a preservation check rather than a universal dominance claim. It tests whether compact FOCI-selected subsets preserve the frozen TransMIL decision and supports the headroom framing: BRCA shows a clear adaptive-$K$ preservation signal (matching the full-bag AUC $0.907$), whereas NSCLC and PANDA expose selection-saturation regimes where random subsets already preserve much of the prediction, which leaves little operating margin for any external selector. Thus, selected-only AUC serves as a triangulation check rather than the primary evidence of FOCI superiority: it reveals when a dataset/backbone pair has enough operating margin for a learned selector to improve over random compact subsets.

\subsection{Deletion-based perturbation faithfulness}
\label{subsec:faithfulness}

Deletion-based perturbation~\cite{petsiuk2018rise,samek2017evaluating} complements SRP by asking whether top-ranked tiles are load-bearing for the model output when removed, rather than how quickly confidence is recovered when they are inserted. Attention-pooling methods often score strongly on this metric because their ranking is part of the aggregation mechanism itself. On TransMIL, where the native ranking is only a CLS-dot-product proxy, FOCI increases NSCLC deletion-AUC from $0.0003$ to $0.0274$, indicating that the readout extracts a more load-bearing ranking than the frozen proxy. Cross-dataset deletion-AUC results are reported in Appendix~\ref{app:deletion_details}.

We therefore treat SRP, deletion, and selected-only AUC as complementary rather than interchangeable rationale-quality axes: SRP measures insertion sufficiency, deletion measures load-bearing removal, and selected-only AUC measures downstream prediction preservation under masked input. No single ranking dominates all three, which is why we use deletion as a faithfulness check rather than as the sole explanation metric.

\section{Conclusion}
\label{sec:conclusion}
FOCI is a post-hoc rationale-highlighting layer for frozen WSI-MIL classifiers: the full-bag prediction is preserved while FOCI selects a compact, output-consistent tile subset that recovers it, evaluated through SRP and the Selection Headroom Index. Across three benchmarks and seven backbones, compact rationales are selection-headroom dependent — TransMIL and ACMIL admit consistent compression, attention-pooling backbones saturate on near-minimal baselines, and hard-selection backbones conflict with an external readout. High slide-level AUC does not by itself imply that a frozen MIL classifier admits a compact rationale; SHI motivates a selection-headroom audit before treating selected tiles as faithful explanations.

\paragraph{Limitations.}
We measure model-output sufficiency only: selected tiles are candidate rationales for the frozen classifier, not annotation-validated clinical evidence. Our experiments use UNI2-h features and binary WSI tasks, so broader encoder evaluation, multiclass settings, external clinical cohorts, and multi-reader validation remain future work; details are in Appendix~\ref{app:limitations}.

\small
\bibliographystyle{unsrtnat}
\bibliography{neurips_2026}






\appendix

\section{Qualitative Illustration}
\label{app:qualitative}

\begin{figure}[!t]
    \centering
    \includegraphics[width=0.95\textwidth]{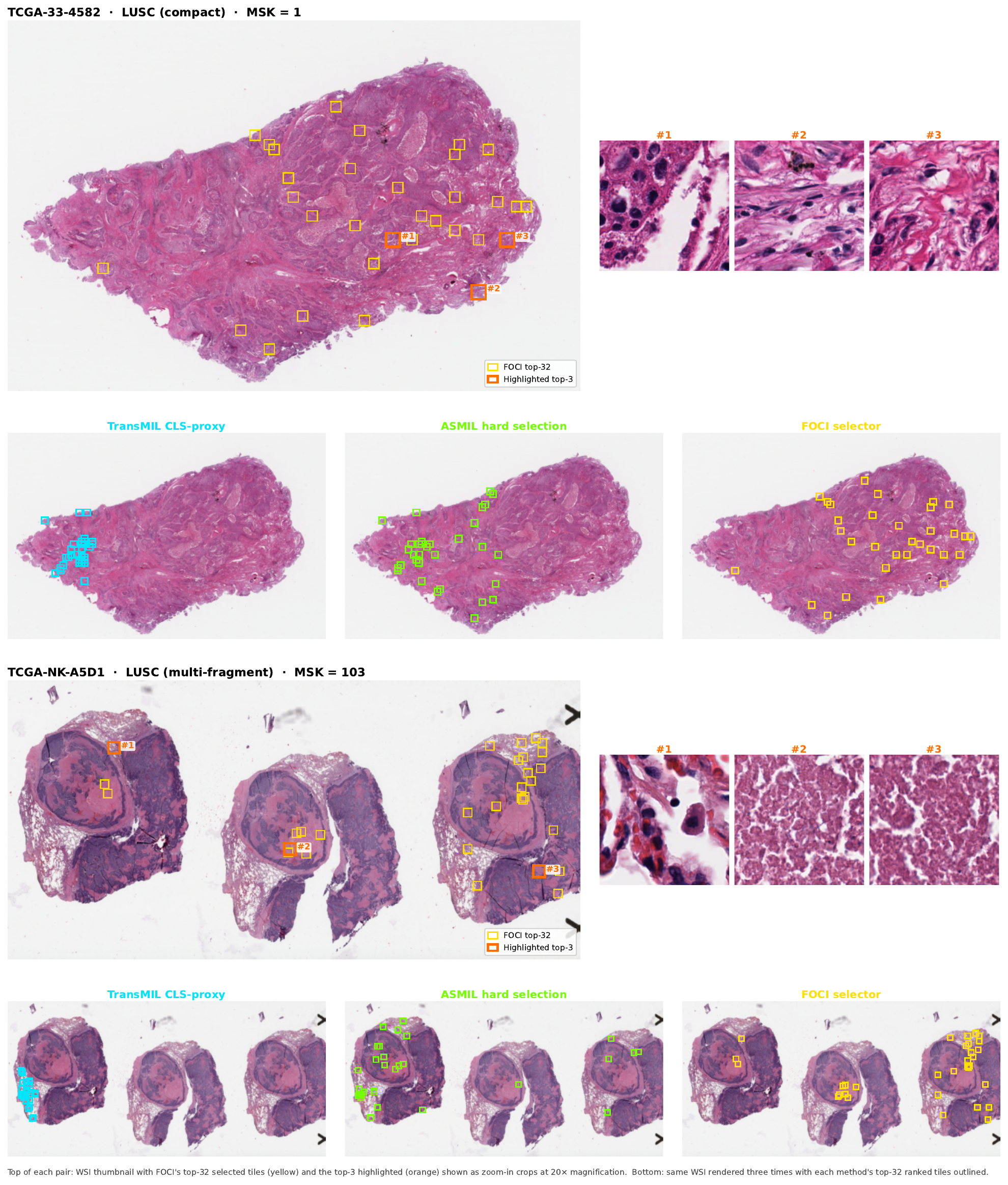}
    \caption{
        \textbf{Qualitative illustration of FOCI selections on two LUSC slides.}
        Each slide is shown twice. \emph{Top row of each pair}: WSI thumbnail with FOCI's top-32 selected tiles outlined in yellow and the top-3 highlighted in orange (\#1, \#2, \#3), plus three zoom-in crops at $20\times$ magnification.
        \emph{Bottom row of each pair}: same WSI rendered three times with each method's top-32 ranked tiles outlined; cyan = TransMIL CLS-proxy ranking, lime = ASMIL hard selection, yellow = FOCI selector.
        \emph{Top slide} (TCGA-33-4582, MSK${=}1$, compact): all three methods cluster in similar regions, consistent with NSCLC's selection-saturation regime.
        \emph{Bottom slide} (TCGA-NK-A5D1, MSK${=}103$, multi-fragment): FOCI concentrates on a single tissue fragment while attention/hard-selection rankings spread across fragments.
        These examples are illustrative and do not establish clinical sufficiency.
    }
    \label{fig:explainability}
\end{figure}

This appendix shows where FOCI-selected tiles appear, in WSI context, relative to two attention/selection baselines on the same input bag. The figure is illustrative and not a claim of clinical sufficiency. Informal pathologist feedback suggested that isolated patch-only review is not aligned with clinical slide review; we therefore present selected tiles only in WSI context. Figure~\ref{fig:explainability} compares FOCI against the TransMIL CLS-proxy ranking and ASMIL hard-selection ranking on two LUSC slides from the TCGA-NSCLC test set; each ranking uses its own scoring source, with FOCI and the TransMIL proxy sharing the frozen TransMIL backbone and ASMIL using its native model.

\textbf{TCGA-33-4582} (compact case, MSK${=}1$): a single patch is enough to cross 90\% confidence. The top row shows FOCI's top-32 selections concentrated in a small densely cellular region, with three highlighted zoom-ins. The bottom row contrasts FOCI's selection against the TransMIL CLS-proxy ranking and ASMIL hard selection on the same bag; all three methods cluster in similar regions, consistent with NSCLC's selection-saturation regime where many tile subsets recover the model's confidence.

\textbf{TCGA-NK-A5D1} (multi-fragment case, MSK${=}103$): the tissue spans multiple fragments with considerable morphological variation, and the model requires 103 patches before crossing the confidence threshold. Here the three methods diverge: FOCI concentrates on a single tissue fragment while attention and hard-selection rankings spread across multiple fragments. The high MSK reflects the slide's genuine complexity rather than a failure of any single ranker.

\textit{Informal visual inspection note.} On informal visual inspection with WSI context, a subset of highlighted FOCI regions appeared plausibly compatible with squamous histology in WSI context. This qualitative review was not blinded, was not systematic, was performed on $n{=}2$ slides for illustration only, and is reported as informal feedback rather than a reader study.

These two cases illustrate that compact low-MSK selections and diffuse high-MSK selections can correspond to visibly different tissue patterns, and that FOCI's selections can differ from attention-based or hard-selection baselines on the same input. They do not establish clinical sufficiency.

\FloatBarrier

\section{Classification Preservation}
\label{app:auc}

FOCI freezes the encoder and MIL backbone and trains only the lightweight selector. Therefore, under standard full-bag inference without masking, the FOCI-augmented model produces the same logits as the standalone backbone. Slide-level AUC is preserved \emph{by construction} for the primary full-bag prediction; FOCI changes only the post-hoc tile ranking and masked keep/drop evaluations. We verified this empirically across all configurations, which confirms full-bag AUC equivalence to four decimal places in every case (7~backbones $\times$ 3~datasets $\times$ 3~seeds).

\section{Per-method SRP ranking-score extraction}
\label{app:ranking_scores}

Each baseline contributes a per-tile ranking score to SRP. Table~\ref{tab:ranking_score_extraction} lists the score used for every method in the 7-backbone matrix. All scores are computed from the trained model on the same pre-filtered bag (top $n_{\text{cap}}{=}1024$ tokens by feature L2 norm; see Appendix~\ref{app:ncap}); a higher score means earlier reveal. TransMIL does not expose a native attention head, so we use a post-encoder CLS-dot-product score $\langle h_{\mathrm{tok}_i}, h_{\mathrm{cls}}\rangle$ as a documented CLS-proxy ranking. For the other six backbones, the ranking score is the model's own pre-softmax attention logit or attribute-attention aggregate, exposed through an \texttt{attn\_logits} interface. FOCI ranks tiles by its selector head, $a_i = \mathrm{MLP}(x_i)$.

\begin{table}[t]
\centering
\caption{Per-tile ranking scores used by each method under SRP. All scores are extracted from the trained model on the same pre-filtered bag and produce a per-tile real number; higher scores are revealed earlier.}
\label{tab:ranking_score_extraction}
\small
\begin{tabular}{p{0.18\textwidth}p{0.74\textwidth}}
\toprule
Method & SRP ranking score \\
\midrule
TransMIL~\cite{shao2021transmil} & Post-encoder CLS-dot-product score $\langle h_{\mathrm{tok}_i}, h_{\mathrm{cls}}\rangle$ (documented CLS-proxy; sole TransMIL ranking source). \\
ABMIL~\cite{ilse2018attention} & Pre-softmax gated-attention logits $\mathbf{w}^{\!\top}(\tanh(\mathbf{V}x_i)\odot \sigma(\mathbf{U}x_i))$. \\
CLAM-SB~\cite{lu2021data} & Pre-softmax gated-attention logits (same gated form as ABMIL); class-specific branch shared with the predicted class. \\
ACMIL~\cite{zhang2024attention} & Mean over the multiple attention branches of the per-tile branch-attention logits. \\
AttriMIL~\cite{CAI2025103631} & Per-class attribute score $\mathrm{inst}_c(x_i)\cdot \exp(a^{\mathrm{logit}}_c(x_i))$, taken as the maximum over classes. \\
ASMIL~\cite{ye2026asmil} & Online model's pre-softmax gated-attention logits; the anchor-EMA path is used only during training. \\
MHIM-MIL~\cite{Tang_2023_ICCV} & Last-layer encoder attention logits exposed through the wrapper's \texttt{return\_attn=True} forward pass. \\
\midrule
FOCI (ours) & Lightweight selector head $a_i = \mathrm{MLP}(x_i)$ trained with sufficiency, exclusion, and contiguity losses. \\
\bottomrule
\end{tabular}
\end{table}

This table is intentionally explicit because SRP is sensitive to ranking quality. The same masking and reveal procedure is applied regardless of how each score is obtained. Methods that perform hard instance selection during training, such as ASMIL and MHIM-MIL, still expose continuous attention logits through this interface; these logits are what we rank for side-by-side SRP comparison in Tables~\ref{tab:main_nsclc}--\ref{tab:main_panda}.

\section{Equal-Budget Comparison}
\label{app:kmax32}

Table~\ref{tab:kmax32} evaluates all methods under a fixed reveal budget of $K_{\max}=32$, which matches the FOCI-STE training target. This equal-budget setting provides a complementary view to the main SRP results at $K_{\max}=256$ (Tables~\ref{tab:main_nsclc}--\ref{tab:main_panda}) by isolating early-ranking quality under a small tile budget. The same headroom-vs-saturation pattern remains visible: FOCI helps when the baseline ranking has room to compress, while near-minimal attention-pooling baselines leave little margin for further reduction.

\begin{table}[t]
\centering
\caption{Equal $K$-budget comparison at $K_{\max}=32$ ($\kappa=0.9$, $n_\mathrm{cap}=1024$, 3-seed mean$\pm$std).
Paired rows show each backbone before and after attaching the FOCI selector. All methods are evaluated under SRP with the same reveal budget, which matches FOCI-STE's training target.
\textbf{Bold}: best per column. \underline{Underline}: second best.}
\label{tab:kmax32}
\resizebox{\textwidth}{!}{%
\renewcommand{\arraystretch}{0.85}
\begin{tabular}{lccccccccc}
\toprule

& \multicolumn{3}{c}{NSCLC} & \multicolumn{3}{c}{BRCA} & \multicolumn{3}{c}{PANDA} \\
\cmidrule(lr){2-4} \cmidrule(lr){5-7} \cmidrule(lr){8-10}
Method & MSK$_{\mathrm{cond}}$ $\downarrow$ & Reach (\%) $\uparrow$ & AUKC $\uparrow$ & MSK$_{\mathrm{cond}}$ $\downarrow$ & Reach (\%) $\uparrow$ & AUKC $\uparrow$ & MSK$_{\mathrm{cond}}$ $\downarrow$ & Reach (\%) $\uparrow$ & AUKC $\uparrow$ \\
\midrule
\multicolumn{10}{l}{\textit{Soft-aggregation backbones}} \\[2pt]
TransMIL         & $2.1\pm0.4$ & $87.7\pm0.5$ & $0.829\pm0.006$
                 & $1.4\pm0.3$ & $81.3\pm4.3$ & $0.809\pm0.011$
                 & $3.1\pm0.2$ & $81.0\pm2.4$ & $0.767\pm0.027$ \\
~~~+~FOCI & $1.7\pm0.1$ & $88.5\pm3.8$ & $0.846\pm0.016$
                 & $1.6\pm0.4$ & $83.7\pm3.3$ & $0.822\pm0.011$
                 & $2.6\pm0.6$ & $84.3\pm3.9$ & $0.824\pm0.033$ \\
\addlinespace
ABMIL            & $\underline{1.2\pm0.0}$ & $91.1\pm0.5$ & $0.891\pm0.005$
                 & $\underline{1.1\pm0.1}$ & $85.4\pm2.6$ & $0.836\pm0.021$
                 & $\underline{1.5\pm0.2}$ & $88.1\pm3.2$ & $\underline{0.897\pm0.017}$ \\
~~~+~FOCI & $\underline{1.2\pm0.1}$ & $\underline{92.8\pm1.8}$ & $\underline{0.899\pm0.009}$
                 & $1.2\pm0.1$ & $84.5\pm4.9$ & $0.826\pm0.027$
                 & $1.7\pm0.7$ & $90.9\pm3.0$ & $0.873\pm0.050$ \\
\addlinespace
CLAM-SB          & $1.3\pm0.1$ & $90.9\pm1.0$ & $0.889\pm0.007$
                 & $\mathbf{1.0\pm0.0}$ & $87.0\pm1.0$ & $\underline{0.848\pm0.003}$
                 & $1.6\pm0.2$ & $88.2\pm1.6$ & $\underline{0.897\pm0.009}$ \\
~~~+~FOCI & $1.3\pm0.3$ & $92.5\pm1.0$ & $0.889\pm0.013$
                 & $1.8\pm0.6$ & $87.7\pm2.9$ & $0.832\pm0.028$
                 & $1.6\pm0.6$ & $91.5\pm2.1$ & $0.875\pm0.039$ \\
\addlinespace
AttriMIL         & $\underline{1.2\pm0.1}$ & $90.4\pm1.4$ & $0.882\pm0.012$
                 & $1.4\pm0.3$ & $83.6\pm3.1$ & $0.818\pm0.022$
                 & $\mathbf{1.3\pm0.1}$ & $90.0\pm1.7$ & $\underline{0.897\pm0.006}$ \\
~~~+~FOCI & $1.4\pm0.2$ & $91.4\pm0.4$ & $0.886\pm0.010$
                 & $1.8\pm0.4$ & $78.9\pm4.6$ & $0.809\pm0.028$
                 & $1.7\pm0.2$ & $91.8\pm0.4$ & $0.877\pm0.007$ \\
\addlinespace
ACMIL            & $1.8\pm0.1$ & $88.8\pm3.0$ & $0.852\pm0.020$
                 & $1.3\pm0.1$ & $\underline{88.5\pm1.5}$ & $0.840\pm0.004$
                 & $1.8\pm0.0$ & $\mathbf{93.8\pm0.4}$ & $0.881\pm0.004$ \\
~~~+~FOCI & $\mathbf{1.1\pm0.1}$ & $90.9\pm0.7$ & $0.887\pm0.003$
                 & $1.3\pm0.1$ & $\mathbf{88.8\pm1.6}$ & $\mathbf{0.856\pm0.005}$
                 & $\underline{1.5\pm0.2}$ & $\underline{93.2\pm1.8}$ & $\mathbf{0.905\pm0.010}$ \\
\midrule
\multicolumn{10}{l}{\textit{Hard-selection backbones}} \\[2pt]
ASMIL            & $\underline{1.2\pm0.1}$ & $\mathbf{93.8\pm1.0}$ & $\mathbf{0.908\pm0.003}$
                 & $2.4\pm0.5$ & $70.9\pm12.0$ & $0.703\pm0.082$
                 & $1.7\pm0.1$ & $82.3\pm1.8$ & $0.817\pm0.014$ \\
~~~+~FOCI & $1.4\pm0.3$ & $87.7\pm1.9$ & $0.840\pm0.025$
                 & $2.3\pm0.7$ & $72.9\pm8.9$ & $0.734\pm0.069$
                 & $3.4\pm1.1$ & $70.6\pm14.8$ & $0.671\pm0.125$ \\
\addlinespace
MHIM-MIL         & $1.5\pm0.0$ & $83.1\pm4.0$ & $0.843\pm0.015$
                 & $2.2\pm0.6$ & $55.9\pm5.7$ & $0.758\pm0.020$
                 & $1.7\pm0.2$ & $88.0\pm1.7$ & $0.890\pm0.009$ \\
~~~+~FOCI & $1.5\pm0.3$ & $81.8\pm2.9$ & $0.827\pm0.006$
                 & $3.0\pm0.7$ & $52.5\pm8.9$ & $0.755\pm0.022$
                 & $\underline{1.5\pm0.0}$ & $87.8\pm2.1$ & $0.884\pm0.009$ \\
\bottomrule
\end{tabular}
}
\end{table}

\section{Sensitivity Analysis}
\label{app:sensitivity}

We analyze the sensitivity of FOCI-STE to hyperparameters that govern the SRP evaluation protocol and the selector budget.

\subsection{Operating confidence threshold $\kappa$}
\label{app:kappa}

The operating confidence threshold $\kappa$ determines when a classifier is deemed sufficiently confident during sequential reveal. Table~\ref{tab:kappa_sensitivity} reports MSK, Reach, and AUKC for FOCI-STE at $\kappa \in \{0.7, 0.8, 0.9, 0.95\}$ across all three datasets. AUKC is invariant to $\kappa$ by construction: it is computed from the full reveal-probability curve, so the threshold affects only which slides reach the operating point and how many tiles are counted toward MSK. Reach generally decreases as $\kappa$ rises. MSK$_\mathrm{cond}$ can shift non-monotonically because it is averaged only over reachable slides: higher thresholds require more tiles for reachable slides, while the hardest slides may drop out of the conditional set. The default $\kappa{=}0.9$ provides a conservative operating point for the main results.

\begin{table}[t]
\centering
\caption{Sensitivity to operating confidence threshold $\kappa$ on FOCI-STE. All results use $K_{\max}=256$ and $n_{\mathrm{cap}}=1024$. Bold $\kappa$: default used in all main results. AUKC is unthresholded and therefore invariant to $\kappa$ for a fixed reveal curve.}
\label{tab:kappa_sensitivity}
\resizebox{\textwidth}{!}{%
\renewcommand{\arraystretch}{0.85}
\begin{tabular}{lccccccccc}
\toprule
& \multicolumn{3}{c}{NSCLC} & \multicolumn{3}{c}{BRCA} & \multicolumn{3}{c}{PANDA} \\
\cmidrule(lr){2-4} \cmidrule(lr){5-7} \cmidrule(lr){8-10}
$\kappa$ & MSK$_{\mathrm{cond}}$ $\downarrow$ & Reach (\%) $\uparrow$ & AUKC $\uparrow$ &
MSK$_{\mathrm{cond}}$ $\downarrow$ & Reach (\%) $\uparrow$ & AUKC $\uparrow$ &
MSK$_{\mathrm{cond}}$ $\downarrow$ & Reach (\%) $\uparrow$ & AUKC $\uparrow$ \\
\midrule
0.70 & $3.68\pm0.59$ & $94.4\pm1.6$ & $0.893\pm0.016$ & $3.68\pm1.28$ & $89.8\pm1.1$ & $0.856\pm0.013$ & $9.01\pm2.79$ & $96.3\pm0.7$ & $0.903\pm0.018$ \\
0.80 & $3.82\pm0.84$ & $94.1\pm1.6$ & $0.893\pm0.016$ & $3.29\pm0.96$ & $88.8\pm1.1$ & $0.856\pm0.013$ & $9.57\pm3.21$ & $95.2\pm0.6$ & $0.903\pm0.018$ \\
\textbf{0.90} & $3.21\pm0.38$ & $90.1\pm4.3$ & $0.893\pm0.016$ & $3.86\pm0.95$ & $85.7\pm2.5$ & $0.856\pm0.013$ & $10.62\pm3.45$ & $92.0\pm1.4$ & $0.903\pm0.018$ \\
0.95 & $2.96\pm1.27$ & $75.8\pm24.2$ & $0.893\pm0.016$ & $5.13\pm0.86$ & $77.0\pm10.3$ & $0.856\pm0.013$ & $11.25\pm3.87$ & $88.8\pm2.0$ & $0.903\pm0.018$ \\
\bottomrule
\end{tabular}
}
\end{table}

\subsection{Pre-filter budget $n_{\mathrm{cap}}$}
\label{app:ncap}

The pre-filter budget $n_{\mathrm{cap}}$ controls how many patches are retained by L2-norm pre-filtering before FOCI re-ranks them. Table~\ref{tab:ncap_sensitivity} varies $n_{\mathrm{cap}}\in\{256,512,1024,2048\}$ while keeping the SRP threshold fixed at $\kappa=0.9$. AUKC varies by less than $0.006$ across all settings and datasets, which indicates that the FOCI-STE ranking is not sharply sensitive to this pre-filter budget. MSK$_\mathrm{cond}$ can shift, especially on BRCA, because changing the candidate pool alters the marginal patch distribution even when the overall ranking signal remains stable.

\begin{table}[t]
\centering
\small
\caption{Sensitivity to pre-filter budget $n_{\mathrm{cap}}$ on FOCI-STE. All results use $\kappa=0.9$ and $K_{\max}=256$. Bold $n_{\mathrm{cap}}$: default.}
\label{tab:ncap_sensitivity}
\resizebox{\textwidth}{!}{%
\renewcommand{\arraystretch}{1.0}
\begin{tabular}{lccccccccc}
\toprule
& \multicolumn{3}{c}{NSCLC} & \multicolumn{3}{c}{BRCA} & \multicolumn{3}{c}{PANDA} \\
\cmidrule(lr){2-4} \cmidrule(lr){5-7} \cmidrule(lr){8-10}
$n_{\mathrm{cap}}$ & MSK$_{\mathrm{cond}}$ $\downarrow$ & Reach (\%) $\uparrow$ & AUKC $\uparrow$ &
MSK$_{\mathrm{cond}}$ $\downarrow$ & Reach (\%) $\uparrow$ & AUKC $\uparrow$ &
MSK$_{\mathrm{cond}}$ $\downarrow$ & Reach (\%) $\uparrow$ & AUKC $\uparrow$ \\
\midrule
256           & $3.24 \pm 1.42$ & $91.1 \pm 2.2$ & $0.892 \pm 0.012$
              & $2.36 \pm 0.86$ & $84.9 \pm 2.4$ & $0.850 \pm 0.012$
              & $8.98 \pm 3.45$ & $92.2 \pm 0.8$ & $0.908 \pm 0.015$ \\
512           & $4.16 \pm 1.36$ & $91.4 \pm 2.2$ & $0.893 \pm 0.010$
              & $2.72 \pm 1.27$ & $85.1 \pm 2.1$ & $0.853 \pm 0.010$
              & $10.60 \pm 3.47$ & $92.3 \pm 1.2$ & $0.905 \pm 0.017$ \\
\textbf{1024} & $3.21 \pm 0.38$ & $90.1 \pm 4.3$ & $0.893 \pm 0.016$
              & $3.86 \pm 0.95$ & $85.7 \pm 2.5$ & $0.856 \pm 0.013$
              & $10.62 \pm 3.45$ & $92.0 \pm 1.4$ & $0.903 \pm 0.018$ \\
2048          & $4.16 \pm 1.42$ & $90.0 \pm 4.9$ & $0.889 \pm 0.019$
              & $5.29 \pm 2.16$ & $86.2 \pm 2.5$ & $0.856 \pm 0.011$
              & $10.69 \pm 3.47$ & $92.1 \pm 1.4$ & $0.903 \pm 0.018$ \\
\bottomrule
\end{tabular}
}
\end{table}

\subsection{Adaptive $K$ training schedule ($\alpha$, $K_{\min}$)}
\label{app:adaptive_k}

Section~\ref{subsec:selected_only} evaluates a separate adaptive-$K$ training schedule,
$K_s=\max(K_{\min},\lfloor \alpha N_s^{\mathrm{real}}\rfloor)$, for the selected-only downstream analysis. This schedule is separate from the fixed-$K{=}32$ FOCI-STE configuration used in the main SRP experiments; $N_s^{\mathrm{real}}$ is the unpadded token count of slide $s$. The default setting ($\alpha=0.03$, $K_{\min}=16$) corresponds to approximately $K\approx30$ at the pre-filter cap $n_{\mathrm{cap}}=1024$ and uses $K=16$ for short slides.

Table~\ref{tab:adaptive_k_settings} reports validation AUKC for the default adaptive schedule on all three datasets, averaged over three seeds, and a single-seed $\alpha$ sweep on NSCLC. The sweep is intended as a sensitivity check rather than a separate model-selection procedure. AUKC varies by roughly $0.01$--$0.02$ across $\alpha\in\{0.01,0.03,0.05\}$, comparable to seed variation, which suggests that the adaptive rule is not sharply sensitive to the budget coefficient. Eval-time $K$-sensitivity at the default-trained selector is reported in \S\ref{subsec:selected_only}.

\begin{table}[h]
\centering
\caption{Adaptive $K$ training schedule sensitivity. Top block: default $\alpha=0.03$ with $K_{\min}=16$, mean$\pm$std over three seeds. Bottom block: single-seed (seed 42) $\alpha$ sweep on NSCLC for sensitivity characterization.}
\label{tab:adaptive_k_settings}
\small
\begin{tabular}{lccc}
\toprule
Dataset & $\alpha$ & $K_{\min}$ & Validation AUKC \\
\midrule
\multicolumn{4}{l}{\textit{Default schedule, three seeds}} \\
NSCLC & 0.03 & 16 & $0.854 \pm 0.011$ \\
BRCA  & 0.03 & 16 & $0.801 \pm 0.005$ \\
PANDA & 0.03 & 16 & $0.817 \pm 0.027$ \\
\midrule
\multicolumn{4}{l}{\textit{$\alpha$ sweep, NSCLC seed 42 only}} \\
NSCLC & 0.01 & 16 & $0.853$ \\
NSCLC & 0.03 & 16 & $0.845$ \\
NSCLC & 0.05 & 16 & $\mathbf{0.863}$ \\
\bottomrule
\end{tabular}
\end{table}

\clearpage
\begin{landscape}
\section{Extended SRP Curves}
\label{app:srp_extended}

\centering
\includegraphics[
    width=0.92\linewidth,
    height=0.78\textheight,
    keepaspectratio
]{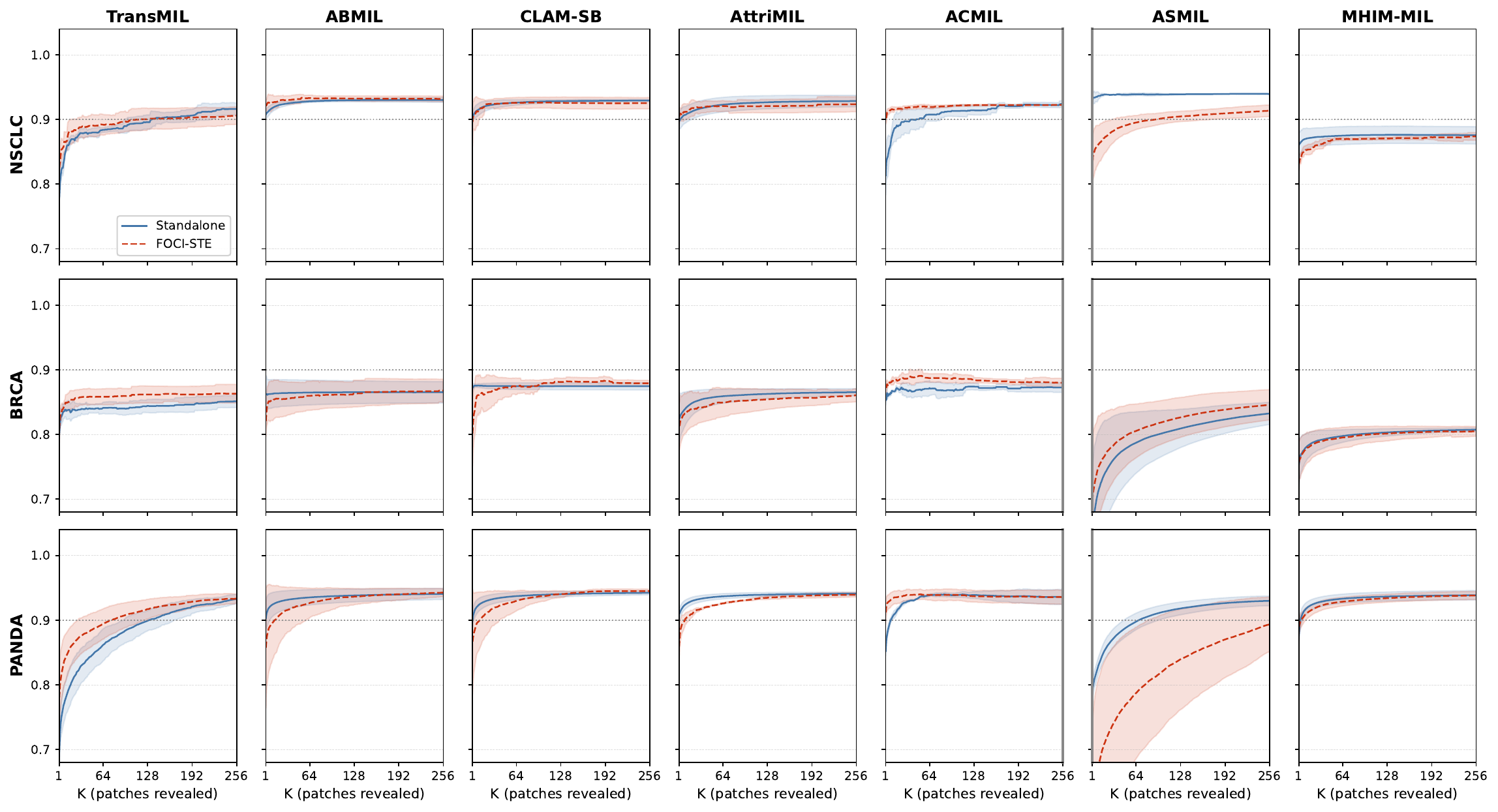}

\vspace{-0.25em}
\captionof{figure}{Extended SRP reveal curves for seven backbones $\pm$ FOCI across NSCLC, BRCA, and PANDA. Curves show the full confidence--$K$ trajectories behind Tables~\ref{tab:main_nsclc}--\ref{tab:main_panda}; blue solid = standalone backbone, red dashed = FOCI-augmented, shaded = $\pm1$ std over three seeds, dotted line = $\kappa=0.9$.}
\label{fig:srp_extended}

\end{landscape}
\clearpage


\section{Additional analyses}
\label{app:additional_analyses}

\subsection{Cross-method SRP reveal curves}
\label{app:srp_curves}

\begin{figure}[!htbp]
    \centering
    \includegraphics[width=\textwidth]{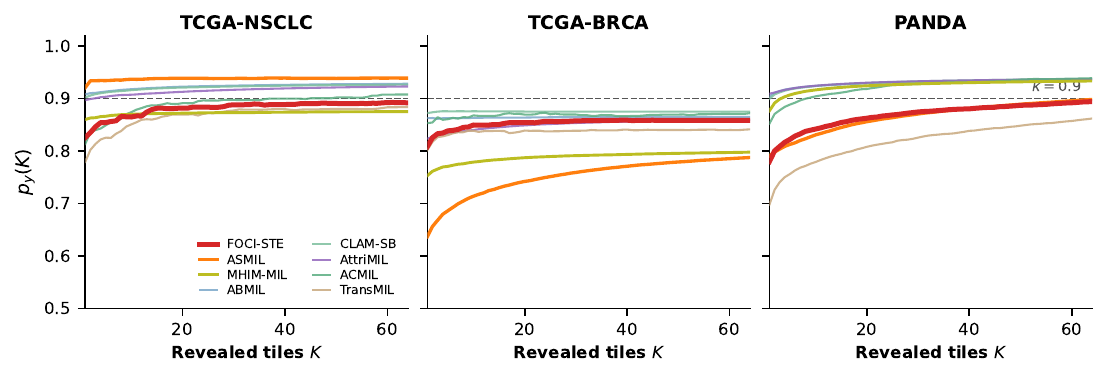}
    \caption{Cross-method SRP confidence curves on TCGA-NSCLC, TCGA-BRCA, and PANDA. True-class probability $p_y(K)$ is averaged over test slides and three seeds as tiles are revealed in descending score order.}
    \label{fig:srp_curve}
\end{figure}

Figure~\ref{fig:srp_curve} shows cross-method SRP confidence curves on TCGA-NSCLC, TCGA-BRCA, and PANDA. True-class probability $p_y(K)$ is averaged over test slides and three seeds as tiles are revealed in descending score order. SRP applies uniformly to any method's tile ranking. ASMIL ranks strongly on NSCLC but collapses on BRCA, a failure mode not visible from slide-level AUC alone. FOCI-STE improves the TransMIL SRP footprint across datasets without the cross-dataset collapse observed in some hard-selection backbones.

\FloatBarrier

\subsection{Inference efficiency}
\label{subsec:efficiency}

FOCI adds only a lightweight selector on top of the frozen backbone. Table~\ref{tab:efficiency} reports standard full-bag inference latency and memory, together with the offline SRP evaluation cost. Adding FOCI introduces negligible measured overhead for standard inference at the reported precision; SRP is a separate offline analysis which requires repeated masked forward passes.

\begin{table}[t]
\centering
\caption{Inference efficiency on NSCLC (mean over 20 slides, RTX 6000 Ada). Paired rows show each backbone before and after attaching FOCI. Standard inference uses the full bag once; SRP columns report the one-time offline cost of sequential reveal evaluation.}
\label{tab:efficiency}
\resizebox{0.85\textwidth}{!}{%
\setlength{\tabcolsep}{4pt}
\begin{tabular}{lrrrrrr}
\toprule
Method & Params & Infer (ms) & $\Delta$ Infer & VRAM (MB) & SRP (ms) & SRP VRAM (MB) \\
\midrule
TransMIL             & 13.5M &  1.4 & \textendash &  264 & 127.2 & 3{,}841 \\
~~~+~FOCI-STE        & 13.7M &  1.4 & \textbf{+0.0} & 263 & 128.1 & 3{,}841 \\
\addlinespace
ABMIL                & 0.49M &  0.6 & \textendash &  176 &   2.1 &    559  \\
~~~+~FOCI-STE        & 0.62M &  0.6 & \textbf{+0.0} & 177 &   2.1 &    559  \\
\addlinespace
ACMIL                & 0.92M &  1.1 & \textendash &  178 & 107.6 &    179  \\
~~~+~FOCI-STE        & 1.05M &  1.1 & \textbf{+0.0} & 179 & 107.6 &    179  \\
\midrule
ASMIL                & 0.56M &  0.4 & \textendash &  177 &   2.1 &    560  \\
CLAM-SB              & 0.49M &  0.4 & \textendash &  176 &   2.1 &    559  \\
MHIM-MIL             & 1.18M &  0.3 & \textendash &  184 &   3.3 &    820  \\
AttriMIL             & 1.71M &  0.7 & \textendash &  185 &  13.8 & 1{,}209 \\
\bottomrule
\end{tabular}%
}
\end{table}

\FloatBarrier

\subsection{Selected-only downstream performance}
\label{subsec:selected_only}

SRP and deletion-based faithfulness characterize rationale ranking quality, but they do not directly answer an audit-relevant question: \emph{if the model only sees the top-$K$ selected tiles, does the slide-level prediction still hold?} Because the backbone is frozen, the primary full-bag prediction is preserved by construction; the meaningful test is whether the selected $K$-tile subset alone preserves the prediction. Table~\ref{tab:selected_only} reports top-$K$ classification metrics in which each (predictor, ranking) pipeline is restricted to its top-$K$ tiles via \texttt{key\_padding\_mask} exclusion of the rest.

\begin{table}[t]
\centering
\caption{Selected-only downstream performance. Each row is a (predictor, ranking) pipeline restricted to top-$K$ tiles. Adaptive $K$ uses $K_s = \max(K_{\min}{=}16, \lfloor\alpha N_s^{\mathrm{real}}\rfloor)$ with $\alpha=0.03$ (average $K\!\approx\!30$ at $n_{\mathrm{cap}}{=}1024$). Random $K{=}32$ uses TransMIL as the predictor for a random-control baseline. ABMIL is reported under its own native backbone and attention ranking.}
\label{tab:selected_only}
\small
\begin{tabular}{llcc}
\toprule
Dataset & Method (Predictor + Ranking) & Top-$K$ AUC & $\Delta$ vs.\ full-bag \\
\midrule
\multirow{5}{*}{NSCLC} & TransMIL (full bag, $K{=}1024$) & $0.974\pm0.003$ & (baseline) \\
                       & TransMIL + FOCI (fixed $K{=}32$) & $0.954\pm0.017$ & $-0.020$ \\
                       & TransMIL + FOCI (adaptive $K_s$, $\bar K{\approx}30$) & $0.963\pm0.004$ & $-0.011$ \\
                       & TransMIL + Random $K{=}32$ & $0.969\pm0.004$ & $-0.005$ \\
                       & ABMIL native (top-$32$) & $0.976\pm0.002$ & (own pipeline) \\
\midrule
\multirow{5}{*}{BRCA}  & TransMIL (full bag, $K{=}1024$) & $0.907\pm0.002$ & (baseline) \\
                       & TransMIL + FOCI (fixed $K{=}32$) & $0.885\pm0.012$ & $-0.022$ \\
                       & TransMIL + FOCI (adaptive $K_s$, $\bar K{\approx}30$) & $0.907\pm0.011$ & $\pm 0.000$ \\
                       & TransMIL + Random $K{=}32$ & $0.881\pm0.007$ & $-0.026$ \\
                       & ABMIL native (top-$32$) & $0.865\pm0.040$ & (own pipeline) \\
\midrule
\multirow{5}{*}{PANDA} & TransMIL (full bag, $K{=}1024$) & $0.989\pm0.002$ & (baseline) \\
                       & TransMIL + FOCI (fixed $K{=}32$) & $0.945\pm0.006$ & $-0.044$ \\
                       & TransMIL + FOCI (adaptive $K_s$, $\bar K{\approx}30$) & $0.934\pm0.012$ & $-0.055$ \\
                       & TransMIL + Random $K{=}32$ & $0.931\pm0.010$ & $-0.058$ \\
                       & ABMIL native (top-$32$) & $0.989\pm0.001$ & (own pipeline) \\
\bottomrule
\end{tabular}
\end{table}

Table~\ref{tab:selected_only} supports the headroom interpretation rather than a universal dominance claim. On BRCA, adaptive-$K$ FOCI preserves the full-bag TransMIL AUC ($0.907$), which outperforms fixed-$K{=}32$ FOCI and random $K{=}32$. On NSCLC and PANDA, random top-$32$ subsets already retain much of the full-bag prediction, which indicates selection-saturation regimes with limited operating margin for any external selector. ABMIL is reported as a native-pipeline reference only, since its backbone and full-bag baseline differ from TransMIL. Adaptive-$K$ robustness on BRCA was further checked by an eval-time sweep over $K\in\{8,16,32,64,128\}$: selected-only AUC reaches $0.906\pm0.015$ even at $K{=}16$, which supports the $K_{\min}{=}16$ floor used in the adaptive rule.

\FloatBarrier

\subsection{Ablation study}
\label{subsec:ablation}

\paragraph{STE vs.\ soft gate.}
Table~\ref{tab:ablation} compares FOCI-STE and FOCI-Soft on NSCLC under the same frozen backbone, losses, and hyperparameters. FOCI-Soft uses a Gumbel-sigmoid relaxation ($T{=}0.5$) with an entropy regularizer ($\lambda_{\text{ent}}{=}0.1$), whereas FOCI-STE uses a hard top-$K$ forward mask with a sigmoid surrogate gradient. FOCI-Soft underperforms the freeze-only control on MSK, consistent with a soft-vs-hard cardinality gap: training uses continuous nonzero gates, whereas SRP evaluates hard top-$K$ subsets. FOCI-STE narrows this mismatch, reduces MSK from $8.03$ to $3.21$, and preserves the same frozen classifier.

\begin{table}[t]
\centering
\caption{Gate formulation ablation on NSCLC ($\kappa{=}0.9$). Values are 3-seed mean except the freeze-only diagnostic control, which uses two seeds.}
\label{tab:ablation}
\resizebox{0.75\columnwidth}{!}{%
\begin{tabular}{lccc}
\toprule
Method & MSK$_{\text{cond}}$ $\downarrow$ & Reach (\%) $\uparrow$ & AUKC $\uparrow$ \\
\midrule
TransMIL (frozen, no selection) & $7.33\pm0.39$ & $91.7\pm0.9$ & $0.890\pm0.008$ \\
Freeze-only (no rationale loss) & $7.60\pm0.18$ & $\mathbf{93.8\pm1.0}$ & $\mathbf{0.906\pm0.006}$ \\
FOCI-Soft (Gumbel gate)         & $8.03\pm0.45$ & $87.1\pm5.2$ & $0.866\pm0.008$ \\
FOCI-STE (hard top-$K$)         & $\mathbf{3.21\pm0.38}$ & $90.1\pm4.3$ & $0.893\pm0.016$ \\
\bottomrule
\end{tabular}}
\end{table}

\paragraph{Frozen vs.\ joint training.}
In a pilot NSCLC/TransMIL run, unfreezing the backbone and training jointly with the rationale losses reduced validation AUC by more than 15 percentage points within two epochs. We therefore freeze the trained MIL backbone and use FOCI as a readout head over a stable feature space rather than as a jointly trained classifier component.

\paragraph{Loss components.}
Table~\ref{tab:lambda_ablation} ablates each loss term by setting it to zero. The main failure mode is Reach collapse: full FOCI-STE reaches $\kappa{=}0.9$ on 90.1\% of slides, whereas removing any single term reduces Reach to 20--34\%. The sufficiency and exclusion terms are both needed to define the keep/drop contrast, and the contiguity term improves stability. Compared with the freeze-only selector, full FOCI-STE trades a small decrease in Reach/AUKC for a large MSK reduction, consistent with optimizing compact rationale recovery rather than uniformly improving every SRP metric.

\begin{table}[t]
\centering
\caption{Loss component ablation on NSCLC ($\kappa{=}0.9$, 3-seed mean$\pm$std). Each row zeroes one loss while keeping the others at their tuned values.}
\label{tab:lambda_ablation}
\resizebox{0.70\columnwidth}{!}{%
\begin{tabular}{lccc}
\toprule
Method & MSK$_{\text{cond}}$ $\downarrow$ & Reach (\%) $\uparrow$ & AUKC $\uparrow$ \\
\midrule
Freeze-only (no rationale loss)   & $7.60\pm0.18$ & $\mathbf{93.8\pm1.0}$ & $\mathbf{0.906\pm0.006}$ \\
Full FOCI-STE                     & $\mathbf{3.21\pm0.38}$ & $90.1\pm4.3$ & $0.893\pm0.016$ \\
\addlinespace
~~~$\lambda_{\text{suff}}{=}0$    & $1.52\pm0.65$ & $34.4\pm14.7$ & $0.519\pm0.012$ \\
~~~$\lambda_{\text{excl}}{=}0$    & $1.78\pm0.65$ & $30.1\pm21.5$ & $0.508\pm0.005$ \\
~~~$\lambda_{\text{contig}}{=}0$  & $27.41\pm21.73$ & $20.3\pm13.7$ & $0.516\pm0.005$ \\
\bottomrule
\end{tabular}}
\end{table}

\FloatBarrier

\section{Per-dataset SRP main result tables}
\label{app:main_tables}

Tables~\ref{tab:main_nsclc}--\ref{tab:main_panda} report per-dataset SRP results at $\kappa{=}0.9$ for each frozen backbone and its FOCI-augmented counterpart. The Selection Headroom Index summary in Table~\ref{tab:shi} is computed from the same per-seed runs. These tables provide the raw per-dataset breakdown behind the family-wise SHI analysis in the main text.

\begin{table}[t]
\centering
\caption{SRP results on \textbf{NSCLC} comparing each backbone with and without FOCI (3-seed mean$\pm$std).
\textbf{Bold}/\underline{underline}: best/second best.
Deltas show change from baseline (\textcolor{red!85!black}{improved}, \textcolor{blue!85!black}{degraded}).}
\label{tab:main_nsclc}
\resizebox{0.68\columnwidth}{!}{%
\renewcommand{\arraystretch}{0.85}
\begin{tabular}{llll}
\toprule
Method & MSK$_\mathrm{cond}$ $\downarrow$ & Reach (\%) $\uparrow$ & AUKC $\uparrow$ \\
\midrule
\multicolumn{4}{l}{\textit{Soft-aggregation backbones}} \\[2pt]
TransMIL         & $7.33\pm0.39$ & $91.7\pm0.9$ & $0.890\pm0.008$ \\
~~~+~FOCI & $3.21\pm0.38$ \dimp{(-4.12)} & $90.1\pm4.3$ \ddeg{(-1.6)} & $0.893\pm0.016$ \dimp{(+.003)} \\
\addlinespace
ABMIL            & $2.65\pm0.73$ & $92.2\pm0.5$ & $0.925\pm0.002$ \\
~~~+~FOCI & $\underline{1.52\pm0.08}$ \dimp{(-1.13)} & $\underline{93.5\pm1.6}$ \dimp{(+1.3)} & $\underline{0.928\pm0.005}$ \dimp{(+.003)} \\
\addlinespace
CLAM-SB          & $2.09\pm0.56$ & $91.7\pm0.5$ & $0.923\pm0.003$ \\
~~~+~FOCI & $1.57\pm0.30$ \dimp{(-0.52)} & $92.7\pm0.8$ \dimp{(+1.0)} & $0.921\pm0.009$ \ddeg{(-.002)} \\
\addlinespace
AttriMIL         & $3.21\pm0.81$ & $92.0\pm1.8$ & $0.920\pm0.010$ \\
~~~+~FOCI & $2.16\pm1.04$ \dimp{(-1.05)} & $92.2\pm0.8$ \dimp{(+0.2)} & $0.917\pm0.010$ \ddeg{(-.003)} \\
\addlinespace
ACMIL            & $6.08\pm1.09$ & $92.4\pm1.4$ & $0.907\pm0.009$ \\
~~~+~FOCI & $1.79\pm0.02$ \dimp{(-4.29)} & $91.5\pm0.8$ \ddeg{(-0.9)} & $0.917\pm0.001$ \dimp{(+.010)} \\
\midrule
\multicolumn{4}{l}{\textit{Hard-selection backbones}} \\[2pt]
ASMIL            & $\mathbf{1.36\pm0.37}$ & $\mathbf{94.1\pm0.6}$ & $\mathbf{0.935\pm0.000}$ \\
~~~+~FOCI & $4.16\pm1.44$ \ddeg{(+2.80)} & $90.9\pm0.7$ \ddeg{(-3.2)} & $0.896\pm0.011$ \ddeg{(-.039)} \\
\addlinespace
MHIM-MIL         & $2.77\pm1.47$ & $84.4\pm3.0$ & $0.872\pm0.014$ \\
~~~+~FOCI & $3.63\pm1.18$ \ddeg{(+0.86)} & $84.2\pm2.2$ \ddeg{(-0.2)} & $0.865\pm0.002$ \ddeg{(-.007)} \\
\bottomrule
\end{tabular}}
\end{table}

\begin{table}[t]
\centering
\caption{SRP results on \textbf{BRCA} comparing each backbone with and without FOCI (3-seed mean$\pm$std).
\textbf{Bold}/\underline{underline}: best/second best.
Deltas show change from baseline (\textcolor{red!85!black}{improved}, \textcolor{blue!85!black}{degraded}).}
\label{tab:main_brca}
\resizebox{0.68\columnwidth}{!}{%
\renewcommand{\arraystretch}{0.85}
\begin{tabular}{llll}
\toprule
Method & MSK$_\mathrm{cond}$ $\downarrow$ & Reach (\%) $\uparrow$ & AUKC $\uparrow$ \\
\midrule
\multicolumn{4}{l}{\textit{Soft-aggregation backbones}} \\[2pt]
TransMIL         & $5.65\pm3.46$ & $84.5\pm2.6$ & $0.840\pm0.010$ \\
~~~+~FOCI & $3.86\pm0.95$ \dimp{(-1.79)} & $85.7\pm2.5$ \dimp{(+1.2)} & $0.856\pm0.013$ \dimp{(+.016)} \\
\addlinespace
ABMIL            & $\mathbf{1.10\pm0.14}$ & $85.4\pm2.6$ & $0.862\pm0.018$ \\
~~~+~FOCI & $3.21\pm1.49$ \ddeg{(+2.11)} & $86.1\pm3.8$ \dimp{(+0.7)} & $0.859\pm0.021$ \ddeg{(-.003)} \\
\addlinespace
CLAM-SB          & $\underline{1.17\pm0.22}$ & $87.1\pm0.9$ & $0.871\pm0.005$ \\
~~~+~FOCI & $4.16\pm3.87$ \ddeg{(+2.99)} & $\underline{90.1\pm1.9}$ \dimp{(+3.0)} & $\underline{0.873\pm0.008}$ \dimp{(+.002)} \\
\addlinespace
AttriMIL         & $3.52\pm1.82$ & $85.7\pm1.6$ & $0.857\pm0.010$ \\
~~~+~FOCI & $6.79\pm3.78$ \ddeg{(+3.27)} & $82.5\pm3.6$ \ddeg{(-3.2)} & $0.849\pm0.017$ \ddeg{(-.008)} \\
\addlinespace
ACMIL            & $3.39\pm0.57$ & $\mathbf{90.3\pm1.1}$ & $0.867\pm0.007$ \\
~~~+~FOCI & $2.25\pm0.55$ \dimp{(-1.14)} & $90.0\pm1.8$ \ddeg{(-0.3)} & $\mathbf{0.880\pm0.005}$ \dimp{(+.013)} \\
\midrule
\multicolumn{4}{l}{\textit{Hard-selection backbones}} \\[2pt]
ASMIL            & $15.83\pm6.63$ & $80.7\pm7.9$ & $0.796\pm0.038$ \\
~~~+~FOCI & $12.81\pm5.96$ \dimp{(-3.02)} & $80.3\pm5.8$ \ddeg{(-0.4)} & $0.814\pm0.037$ \dimp{(+.018)} \\
\addlinespace
MHIM-MIL         & $12.2\pm5.78$ & $60.4\pm3.0$ & $0.797\pm0.007$ \\
~~~+~FOCI & $14.94\pm11.57$ \ddeg{(+2.74)} & $57.8\pm4.8$ \ddeg{(-2.6)} & $0.795\pm0.013$ \ddeg{(-.002)} \\
\bottomrule
\end{tabular}}
\end{table}

\begin{table}[t]
\centering
\caption{SRP results on \textbf{PANDA} comparing each backbone with and without FOCI (3-seed mean$\pm$std).
\textbf{Bold}/\underline{underline}: best/second best.
Deltas show change from baseline (\textcolor{red!85!black}{improved}, \textcolor{blue!85!black}{degraded}).}
\label{tab:main_panda}
\resizebox{0.68\columnwidth}{!}{%
\renewcommand{\arraystretch}{0.85}
\begin{tabular}{llll}
\toprule
Method & MSK$_\mathrm{cond}$ $\downarrow$ & Reach (\%) $\uparrow$ & AUKC $\uparrow$ \\
\midrule
\multicolumn{4}{l}{\textit{Soft-aggregation backbones}} \\[2pt]
TransMIL         & $16.5\pm2.0$ & $93.4\pm1.4$ & $0.881\pm0.017$ \\
~~~+~FOCI & $10.62\pm3.45$ \dimp{(-5.88)} & $92.0\pm1.4$ \ddeg{(-1.4)} & $0.903\pm0.018$ \dimp{(+.022)} \\
\addlinespace
ABMIL            & $4.08\pm1.64$ & $90.5\pm2.0$ & $0.933\pm0.011$ \\
~~~+~FOCI & $5.37\pm4.18$ \ddeg{(+1.29)} & $94.6\pm1.1$ \dimp{(+4.1)} & $0.927\pm0.019$ \ddeg{(-.006)} \\
\addlinespace
CLAM-SB          & $4.89\pm1.46$ & $91.1\pm0.8$ & $\mathbf{0.934\pm0.005}$ \\
~~~+~FOCI & $4.81\pm2.89$ \dimp{(-0.08)} & $\underline{95.0\pm1.5}$ \dimp{(+3.9)} & $0.931\pm0.011$ \ddeg{(-.003)} \\
\addlinespace
AttriMIL         & $\underline{3.25\pm0.87}$ & $91.8\pm1.2$ & $\underline{0.934\pm0.004}$ \\
~~~+~FOCI & $4.35\pm1.18$ \ddeg{(+1.10)} & $94.5\pm1.0$ \dimp{(+2.7)} & $0.926\pm0.001$ \ddeg{(-.008)} \\
\addlinespace
ACMIL            & $4.78\pm0.75$ & $\mathbf{97.2\pm0.1}$ & $0.930\pm0.005$ \\
~~~+~FOCI & $\mathbf{3.09\pm0.50}$ \dimp{(-1.69)} & $\underline{95.0\pm1.5}$ \ddeg{(-2.2)} & $0.933\pm0.009$ \dimp{(+.003)} \\
\midrule
\multicolumn{4}{l}{\textit{Hard-selection backbones}} \\[2pt]
ASMIL            & $11.20\pm0.44$ & $89.6\pm2.7$ & $0.903\pm0.010$ \\
~~~+~FOCI & $27.35\pm17.25$ \ddeg{(+16.15)} & $88.3\pm5.8$ \ddeg{(-1.3)} & $0.817\pm0.081$ \ddeg{(-.086)} \\
\addlinespace
MHIM-MIL         & $5.30\pm0.95$ & $91.6\pm1.1$ & $0.930\pm0.007$ \\
~~~+~FOCI & $6.25\pm1.24$ \ddeg{(+0.95)} & $92.0\pm1.2$ \dimp{(+0.4)} & $0.927\pm0.007$ \ddeg{(-.003)} \\
\bottomrule
\end{tabular}}
\end{table}

\FloatBarrier

\subsection{Brief per-dataset interpretation}
\label{app:per_dataset_discussion}

\paragraph{NSCLC.}
NSCLC shows the clearest positive-headroom pattern among soft-aggregation backbones. FOCI reduces MSK for all five soft-aggregation backbones, including TransMIL ($7.33\to3.21$), ABMIL ($2.65\to1.52$), and ACMIL ($6.08\to1.79$). Hard-selection backbones behave differently: ASMIL has the best native MSK and AUKC on NSCLC, but attaching an external FOCI selector inflates its sufficient set, consistent with an architectural conflict mode.

\paragraph{BRCA.}
BRCA exposes the selection-saturation regime most clearly for attention-pooling backbones. ABMIL and CLAM-SB already reach near-single-tile MSK ($1.10$ and $1.17$), which leaves little room for an external selector; ABMIL+FOCI and CLAM-SB+FOCI therefore inflate MSK despite small AUKC changes. In contrast, TransMIL and ACMIL retain headroom: FOCI reduces TransMIL MSK from $5.65$ to $3.86$, and ACMIL+FOCI achieves the highest BRCA AUKC ($0.880$) with reduced MSK.

\paragraph{PANDA.}
PANDA has higher baseline MSK for several backbones, consistent with a more distributed prediction footprint under the binarized prostate grading task. FOCI reduces TransMIL MSK from $16.5$ to $10.62$ and ACMIL MSK from $4.78$ to $3.09$. However, ABMIL+FOCI and AttriMIL+FOCI are less stable, and ASMIL+FOCI strongly degrades, which again shows that FOCI is a readout probe whose usefulness depends on backbone selection headroom rather than a universal improvement module.

\paragraph{Takeaway.}
Across datasets, the appendix tables support the main SHI result: compact post-hoc rationales are available when the frozen backbone has selection headroom, the regime saturates when the native ranking is already near-minimal, and the readout can conflict with hard-selection backbones. These tables provide the raw per-dataset breakdown for the family-wise summary in Table~\ref{tab:shi}.

\FloatBarrier

\section{Deletion-based perturbation details}
\label{app:deletion_details}

Deletion-based perturbation asks whether top-ranked tiles are load-bearing when removed from the input. This is complementary to SRP insertion: deletion measures removal impact, whereas SRP measures confidence recovery as ranked tiles are inserted. Negative values indicate that deleting the ranked tiles increases the true-class probability on average, which usually reflects saturation or noisy ranking under that perturbation protocol.

\begin{table}[t]
\centering
\caption{Deletion-based perturbation faithfulness on TCGA-NSCLC (3-seed mean$\pm$std). Faithfulness AUC summarizes the drop in true-class probability when the top-$K$ ranked tiles are deleted, averaged over $K \in \{16, 32, 64, 128, 256\}$~\citep{petsiuk2018rise}. Higher values indicate a more load-bearing ranking.}
\label{tab:faithfulness}
\small
\begin{tabular}{lcccccc}
\toprule
Method & Faith.\ AUC $\uparrow$ & $\Delta p_y$@K=16 & @K=32 & @K=64 & @K=128 & @K=256 \\
\midrule
TransMIL              & $0.0003\pm0.0007$ & $-0.000$ & $-0.000$ & $0.000$ & $0.000$ & $0.001$ \\
ASMIL                 & $0.0087\pm0.0013$ & $0.002$  & $0.003$  & $0.006$ & $0.010$ & $0.015$ \\
FOCI (ours, on TransMIL) & $0.0274\pm0.0213$ & $0.008$ & $0.013$ & $0.020$ & $0.029$ & $0.046$ \\
AttriMIL              & $0.0444\pm0.0069$ & $0.004$  & $0.014$  & $0.026$ & $0.046$ & $0.084$ \\
MHIM-MIL              & $0.0537\pm0.0072$ & $0.016$  & $0.023$  & $0.036$ & $0.058$ & $0.090$ \\
CLAM-SB               & $0.0551\pm0.0116$ & $0.010$  & $0.025$  & $0.039$ & $0.061$ & $0.089$ \\
\textbf{ABMIL}        & $\mathbf{0.0736\pm0.0142}$ & $\mathbf{0.023}$ & $\mathbf{0.035}$ & $\mathbf{0.057}$ & $\mathbf{0.080}$ & $\mathbf{0.116}$ \\
\bottomrule
\end{tabular}
\end{table}

\FloatBarrier

\begin{table}[t]
\centering
\caption{Cross-dataset deletion-based faithfulness for the three methods with comparable per-tile scoring (3-seed mean$\pm$std, normalized by $K_{\max}{=}256$ to match Table~\ref{tab:faithfulness}). Higher values indicate a more load-bearing ranking; negative values indicate that deleting the ranked tiles increases the true-class probability on average. Absolute scale differs across datasets because of saturation, so comparisons are within-dataset. \textbf{Bold}: best within each dataset.}
\label{tab:faithfulness_xds}
\small
\begin{tabular}{lccc}
\toprule
Method & NSCLC & BRCA & PANDA \\
\midrule
TransMIL                          & $0.0003\pm0.0007$ & $0.0010\pm0.0010$ & $0.0491\pm0.0126$ \\
ASMIL                             & $0.0087\pm0.0013$ & $-0.0054\pm0.0013$ & $\mathbf{0.1681\pm0.0168}$ \\
\textbf{FOCI (ours, on TransMIL)} & $\mathbf{0.0274\pm0.0213}$ & $\mathbf{0.0097\pm0.0089}$ & $0.0747\pm0.0131$ \\
\bottomrule
\end{tabular}
\end{table}

\FloatBarrier

\section{FOCI-STE: full technical details}
\label{app:ste_details}

\paragraph{Hard top-$K$ with straight-through (FOCI-STE).}
FOCI-Soft uses continuous Concrete gates during training, whereas SRP evaluates hard ranked subsets at test time. FOCI-STE reduces this soft-vs-hard cardinality mismatch by enforcing an exactly $K$-sparse binary mask in the forward pass while routing gradients through a sigmoid surrogate~\cite{bengio2013estimatingpropagatinggradientsstochastic}. Given selector logits $a_{s,i}$ for slide $s$, we define
\begin{equation}
    m_{s,i} = \mathbf{1}[a_{s,i} \in \mathrm{top}\text{-}K(a_s)],
    \qquad
    \sum_i m_{s,i}=K,
\end{equation}
and use the straight-through gate
\begin{equation}
    \tilde{m}_{s,i}
    =
    m_{s,i}
    +
    \sigma(a_{s,i})
    -
    \mathrm{stopgrad}\!\left(\sigma(a_{s,i})\right).
\end{equation}
The forward value of $\tilde{m}_{s,i}$ equals the binary mask $m_{s,i}$, while the backward gradient follows the sigmoid surrogate,
\begin{equation}
    \frac{\partial \tilde{m}_{s,i}}{\partial a_{s,i}}
    =
    \sigma'(a_{s,i}).
\end{equation}
Thus, exactly $K$ tiles are selected in every forward pass, but the selector logits still receive dense surrogate gradients during optimization.

During training, the auxiliary keep/drop views are realized through multiplicative straight-through gating. During SRP evaluation, unrevealed tokens are excluded through the model's masking interface (\texttt{key\_padding\_mask}). These two operations are not pointwise identical for every MIL pooling architecture, but they impose the same hard cardinality constraint and use the same tile ranking. This is the relevant alignment for our audit setting: the selector is trained to produce a compact ordered subset, and SRP evaluates the resulting order under hard reveal.

Although the hard top-$K$ operator fixes the forward-pass cardinality, we retain a small per-bag budget regularizer,
\begin{equation}
    \mathcal{L}_{\mathrm{budget}}
    =
    \sum_i \tilde{m}_{s,i},
    \qquad
    \lambda_{\mathrm{budget}} = 5\times10^{-3}.
\end{equation}
Its forward value is constant at $K$, but its backward pass provides a small stabilizing gradient to the underlying selector scores through $\sigma(a_s)$. This term is therefore not used to enforce sparsity in FOCI-STE (top-$K$ already does that), but to regularize the score scale around the hard selection boundary. The FOCI-Soft counterpart uses continuous gates $z_s$ and is described in \S\ref{subsec:losses}.

FOCI-STE is one parameterization of the same frozen-backbone audit framework as FOCI-Soft. The central object of study is not the gate parameterization itself, but whether a frozen WSI-MIL classifier exhibits selection headroom under a consistent tile ranking.

\section{Loss term details}
\label{app:loss_details}

This appendix clarifies three implementation details that are abbreviated in the main method: the budget regularizer, the FOCI-Soft entropy term, and the shorthand ``sufficiency objective.''

\paragraph{Budget regularizer.}
For FOCI-Soft, the budget term is applied to the continuous Gumbel-sigmoid gates:
\[
\mathcal{L}_{\mathrm{budget}}
=
\sum_i z_{s,i},
\]
where $z_{s,i}\in(0,1)$ is the soft gate for tile $i$ in slide $s$. This term discourages diffuse high-mass gates and encourages the selector to use a compact subset.

For FOCI-STE, the hard top-$K$ forward mask already fixes the selected cardinality exactly, $\sum_i m_{s,i}=K$. We therefore apply the budget term to the straight-through gate $\tilde m_s$ defined in Appendix~\ref{app:ste_details}:
\[
\mathcal{L}_{\mathrm{budget}}
=
\sum_i \tilde m_{s,i}.
\]
Its forward value is constant at $K$, but its backward pass provides a small stabilizing gradient through the sigmoid surrogate. Thus, in FOCI-STE, $\mathcal{L}_{\mathrm{budget}}$ regularizes score scale near the rank-$K$ boundary rather than enforcing sparsity. We use $\lambda_{\mathrm{budget}}=5\times10^{-3}$ in both FOCI-Soft and FOCI-STE.

\paragraph{Sufficiency objective shorthand.}
We use ``sufficiency objective'' as shorthand for the keep-bag terms $\mathcal{L}_{\mathrm{suff}}$ and $\mathcal{L}_{\mathrm{hinge}}$. Both are computed on the keep bag, but they enter the total objective with different weights: $\mathcal{L}_{\mathrm{suff}}$ encourages recovery of the target-class output, while $\mathcal{L}_{\mathrm{hinge}}$ enforces the operating-confidence margin used during selector training.

\paragraph{FOCI-Soft entropy term.}
FOCI-Soft uses continuous gates and therefore does not impose an exact tile count during training. To prevent diffuse fractional masks, we add an entropy penalty
\[
\lambda_{\mathrm{ent}}\mathcal{H}(z_s),
\qquad
\mathcal{H}(z_s)
=
-\frac{1}{N_s}
\sum_i
\left[
z_{s,i}\log z_{s,i}
+
(1-z_{s,i})\log(1-z_{s,i})
\right],
\]
which pushes the continuous gates toward binary values. This entropy term is used only for FOCI-Soft; FOCI-STE obtains exact hard cardinality through the top-$K$ forward mask.

\section{Full experimental setup details}
\label{app:setup_full}

\subsection{Datasets and features}
\label{subsec:datasets}

We evaluate on three public WSI benchmarks (TCGA-NSCLC, TCGA-BRCA, and PANDA), all with slide-level labels and no patch-level annotation.

\paragraph{NSCLC.}
The non-small-cell lung cancer cohort comprises 1{,}043 slides (729 train / 105 validation / 209 test). Each slide is labeled LUAD (lung adenocarcinoma) or LUSC (lung squamous cell carcinoma).

\paragraph{BRCA.}
The breast cancer cohort contains 1{,}126 slides (724 train / 179 validation / 223 test), labeled as invasive ductal carcinoma (IDC) versus other subtypes. The class distribution is skewed ($\approx\!73\%$ IDC), and the boundary between IDC and the rarer subtypes is histologically subtle.

\paragraph{PANDA.}
The PANDA prostate grading dataset~\cite{bulten2022artificial} has 10{,}615 slides (6{,}793 train / 1{,}699 validation / 2{,}123 test). Labels are binarized as benign (ISUP grade 0) versus malignant (ISUP grade $\geq\!1$).

\paragraph{Features.}
Slides are tiled at $20\times$ magnification into 256$\times$256 patches. We extract $d{=}1536$-dimensional features using frozen UNI2-h~\cite{chen2024towards}, a vision transformer pretrained on a large-scale histology corpus. Features are extracted once and stored as HDF5 files, and the encoder is never updated. For both FOCI training and SRP evaluation, slides with more than $n_{\text{cap}}{=}1024$ patches are pre-filtered to the top $n_{\text{cap}}$ tokens by feature L2 norm before MIL aggregation; see Appendix~\ref{app:ncap} for sensitivity.

\subsection{Implementation details}
\label{subsec:impl}

\paragraph{Backbone.}
The primary FOCI-STE backbone is a four-layer TransMIL~\cite{shao2021transmil} with $d_{\text{model}}{=}512$, eight attention heads, and a learned $[\text{CLS}]$ token, pretrained for 20 epochs with cross-entropy on the full bag. For cross-backbone experiments, FOCI is additionally applied post-hoc to ABMIL ($d{=}512$), CLAM-SB ($d{=}256$), AttriMIL ($d{=}512$), ACMIL ($d{=}512$), ASMIL ($d{=}256$), and MHIM-MIL ($d{=}512$), each pretrained independently with its original objective. In all cases, the encoder and MIL backbone are fully frozen during FOCI training.

\paragraph{FOCI-STE training.}
The selection module is a two-layer MLP ($d{=}512 \to 256 \to 1$) with 132{,}609 parameters, which is under 1\% of the primary TransMIL pipeline. We train the selector for 30 epochs with a 5-epoch linear warmup using AdamW with cosine annealing from $10^{-4}$ to $10^{-5}$; the selector uses a 5$\times$ learning-rate multiplier relative to the base schedule and an AdamW weight decay of $0.3$. The frozen encoder and backbone are not optimized. Batch size is 2, and slides are padded to $n_{\text{cap}}$ tokens. FOCI-STE selects exactly $K{=}32$ tiles per slide in the forward pass.

The sufficiency cross-entropy and exclusion losses are weighted equally ($\lambda_{\text{suff}}{=}\lambda_{\text{excl}}{=}0.5$), with the keep-bag confidence hinge added at $\lambda_{\text{hinge}}{=}1.0$ and a light spatial compactness term ($\lambda_{\text{contig}}{=}0.01$) to encourage contiguous selections without dominating the rationale losses. The training sufficiency target is set to $\tau{=}0.9$, numerically matching the SRP operating threshold $\kappa{=}0.9$ used in the main evaluation; the drop-bag tolerance in the exclusion loss (\S\ref{subsec:losses}) is set to $\beta{=}0.2$. For FOCI-Soft, we add budget and entropy penalties ($\lambda_{\text{budget}}{=}5 \times 10^{-3}$, $\lambda_{\text{ent}}{=}0.1$) to push the continuous gates toward binary values.

Unless otherwise stated, FOCI-STE is trained with the fixed $K{=}32$ budget above. The adaptive-$K$ schedule $K_s = \max(16, \lfloor 0.03 N_s^{\mathrm{real}} \rfloor)$ is evaluated separately in \S\ref{subsec:selected_only} for selected-only downstream analysis and in Appendix~\ref{app:adaptive_k} for budget-sensitivity analysis. All reported results are averaged over three seeds and evaluated at $\kappa{=}0.9$.

\subsection{Baselines}
\label{subsec:baselines}

We compare seven MIL methods, all trained with the same frozen UNI2-h features.
\textbf{TransMIL}~\cite{shao2021transmil} is the primary frozen backbone without rationale selection.
\textbf{ABMIL}~\cite{ilse2018attention} uses scalar attention weights.
\textbf{CLAM-SB}~\cite{lu2021data} adds instance-level discrimination via an auxiliary loss.
\textbf{AttriMIL}~\cite{CAI2025103631} decomposes attention across attribute heads.
\textbf{ACMIL}~\cite{zhang2024attention} uses multiple attention branches with masked patch training to capture complementary diagnostic regions.
\textbf{ASMIL}~\cite{ye2026asmil} trains with top-$K$ attention sampling.
\textbf{MHIM-MIL}~\cite{Tang_2023_ICCV} uses masked hard instance mining.

Each baseline is re-evaluated under SRP using its own native per-tile ranking score, summarized in Table~\ref{tab:ranking_score_extraction} (Appendix~\ref{app:ranking_scores}).

\section{Limitations and future work}
\label{app:limitations}

\paragraph{Limitations.}
SRP measures model-output sufficiency, not annotation-validated clinical sufficiency. The selected tiles are candidate rationales for a frozen classifier and do not establish pathologist-level diagnostic sufficiency or clinical utility. Our main experiments use UNI2-h features; because the current FOCI-STE pipeline has not been evaluated across a broad set of pathology encoders, we do not claim universal encoder agnosticism. Extending FOCI to ground-truth tumor annotations (e.g., CAMELYON16/17), broader encoder benchmarks, external clinical cohorts, and multi-reader pathologist studies would be needed to argue clinical relevance beyond model-sufficient rationale highlighting; we leave these directions to future work.


\section{Predicted-class SRP variant (audit-time view)}
\label{app:predicted_class_srp}

The main SRP analysis tracks $p_y(K)$ against the ground-truth label $y$, which jointly assesses confidence recovery and correctness on labeled test sets. In an audit-time setting, however, explanations are often requested for the model's own predicted class $\hat{y}=\arg\max_c f_c(X)$. We therefore report a predicted-class SRP variant using the same insertion-style reveal protocol, but with K-curves tracking $p_{\hat{y}}(K)$.

For the binary classification tasks studied here, this variant can be recovered from the stored K-curves without retraining or re-evaluating any model: for slides where $\hat{y}=y$, $p_{\hat{y}}(K)=p_y(K)$, and for slides where $\hat{y}\ne y$, $p_{\hat{y}}(K)=1-p_y(K)$. We report this audit-time view for the TransMIL baseline and TransMIL+FOCI on all three datasets.

\begin{table}[h]
\centering
\caption{Predicted-class SRP variant on TransMIL baseline vs. TransMIL+FOCI (3-seed mean$\pm$std at $\kappa{=}0.9$). Reach, MSK$_\mathrm{cond}$, and AUKC are computed against the model's own predicted class $\hat{y}$ rather than the ground-truth label $y$. The qualitative pattern matches ground-truth SRP: FOCI compresses MSK on all three datasets, which supports the interpretation that the readout recovers the frozen model's own decision rather than exploiting label-specific evaluation.}
\label{tab:predicted_class_srp}
\small
\setlength{\tabcolsep}{6pt}
\begin{tabular}{lcccc}
\toprule
Dataset & Method & Reach$_{\hat{y}}$ & MSK$^{\hat{y}}_\mathrm{cond}$ $\downarrow$ & AUKC$^{\hat{y}}$ $\uparrow$ \\
\midrule
\multirow{2}{*}{NSCLC} & TransMIL baseline & $0.984$ & $7.55 \pm 0.77$ & $0.956 \pm 0.005$ \\
                       & TransMIL+FOCI     & $0.951$ & $\mathbf{3.52 \pm 0.54}$ & $0.946 \pm 0.034$ \\
\addlinespace
\multirow{2}{*}{BRCA}  & TransMIL baseline & $0.934$ & $6.11 \pm 3.48$ & $0.932 \pm 0.020$ \\
                       & TransMIL+FOCI     & $0.928$ & $\mathbf{4.17 \pm 1.23}$ & $0.936 \pm 0.020$ \\
\addlinespace
\multirow{2}{*}{PANDA} & TransMIL baseline & $0.966$ & $16.82 \pm 1.32$ & $0.914 \pm 0.006$ \\
                       & TransMIL+FOCI     & $0.940$ & $\mathbf{10.81 \pm 3.37}$ & $0.931 \pm 0.015$ \\
\bottomrule
\end{tabular}
\end{table}

Predicted-class SRP changes both the reachable set and the target probability being tracked, so its MSK is not expected to match ground-truth SRP exactly. In our binary tasks, the qualitative compression pattern remains the same: TransMIL+FOCI reduces predicted-class MSK on all three datasets. Reach is generally higher under predicted-class SRP because correctness is no longer required. We report predicted-class SRP as a complementary audit view; ground-truth SRP remains the appropriate benchmark for rationale quality on labelled test sets. For binary classification, this analysis also characterizes recovery of the model's own decision on incorrectly classified slides. Multiclass extensions require tracking $p_{\hat{y}}(K)$ directly and are not addressed in this paper.

\clearpage

\newpage

\end{document}